\documentclass[aps,amsmath,amssymb]{revtex4-1}

\usepackage[pdftex]{graphicx}
\usepackage{graphicx}
\usepackage{bm}
\usepackage{xr}
\usepackage{here}

\usepackage{color}

\newcommand{\argmax}{\mathop{\rm arg~max}\limits}

\def\mean#1{\left< #1 \right>}
\newcommand{\NY}[1]{{\color{black} #1}}
\newcommand{\NYY}[1]{{\color{black} #1}}
\newcommand{\NYYY}[1]{{\color{black} #1}}

\usepackage{amsmath}	
\begin{document}

\preprint{}

\title{
Bayesian Modelling of Pattern Formation from One Snapshot of Pattern
}

\author{Natsuhiko Yoshinaga}
\email[corresponding authors: ]{yoshinaga@tohoku.ac.jp}
\affiliation{
WPI - Advanced Institute for Materials Research, Tohoku University,
Sendai 980-8577, Japan
}
\affiliation{
MathAM-OIL, AIST,
Sendai 980-8577, Japan
}

\author{Satoru Tokuda}
\affiliation{
Research Institute for Information Technology, Kyushu University, Kasuga 816-8580, Japan
}
\affiliation{
MathAM-OIL, AIST,
Sendai 980-8577, Japan
}

   \begin{abstract}
Partial differential equations (PDE) have been widely used to reproduce
    patterns in nature and to give insight into the mechanism underlying
    pattern formation. Although many PDE models have been proposed, they
    rely on the pre-request knowledge of physical laws and symmetries,
    and developing a model to reproduce a given desired pattern remains
    difficult. We propose a novel method, referred to as Bayesian
    modelling of PDE (BM-PDE), to estimate the best \NYYY{dynamical} PDE for one snapshot
    of a target pattern under the stationary state \NY{without ground truth}.
    We apply BM-PDE to nontrivial patterns,
    such as quasi-crystals \NY{(QCs)}, a double gyroid and Frank Kasper
    structures.
    \NY{
By using the estimated parameters for the approximant of QCs, we successfully generate, for the first time,
    three-dimensional dodecagonal QCs from a PDE model.
    }
    Our method works for noisy patterns and the pattern
    synthesised without the ground truth parameters, which are required
    for the application toward experimental data.
   \end{abstract}


\maketitle


The design of structures of materials is one
of the most important
issues in various fields of physical science, as their structures are related to
their physical properties.
The structures are often characterised by periodic or quasi-periodic
order.
These ordered structures, which we call {\it pattern}, are ubiquitous in
nature ranging from fluid
convection\cite{swift:1977} to the microphase separation of block copolymers\cite{bates:1990,Harrison:2000} and atomic and molecular crystals\cite{Elder:2002,Provatas:2011}.
Surprisingly, the same pattern appears in different systems with
completely different length scales\cite{cross:1993}.
Complex patterns such as quasi-crystal (QC), double gyroid (DG) and Frank Kasper (FK) phases
appear not only in metallic alloy\cite{Frank:1959,Shechtman:1984} but also in soft
materials such as block copolymers\cite{bates:1990,Bates:2019}, biomaterials\cite{Scherer:2013}, surfactants\cite{Yue:2016}, liquid crystals\cite{Zeng:2004} and colloidal assemblies\cite{Hynninen:2007}.

 \NY{
Understanding  a
generic mechanism of symmetry selection is an important step to
understand structural pattern formation. 
A continuum approach, using nonlinear partial differential equations
(PDEs), is useful for this
purpose\cite{cross:1993,Provatas:2011}.
For example, it was shown that at least two length scales are necessary
for the formation of QCs \NYY{by using a phenomenological model}\cite{Lifshitz:1997}.}
These studies have clarified generic pictures on
a specific pattern. 
Still, when encountering a new pattern, we need to
 find a governing equation and parameters.
 They require
 a sophisticated guess and trial-and-error.
 \NY{
Therefore, it is necessary to develop a systematic method to estimate the best model for the target
 pattern at hand.
 }

\NY{
Recent developments of imaging techniques give us various structural
information, and thus it is desired to understand pattern formation and
the design principle
 of a desired structure \cite{Kalinin:2015}.
 There are two challenging issues to consider the inverse structural design
 applied to the real-world data.
 First, the structural data is often stationary because a high-resolution
 time-dependent image is hard to acquire.
 Experimentally obtained structure is stable, or at least
 meta-stable.
 We want to estimate the best \NYYY{dynamical} model to reproduce the target pattern as
 a (meta-)stable structure from stationary data.
 Second, there is no ground truth of the target structure; the true model to reproduce the experimental data is not
 available.
 \NYY{
 Still, finding a surrogate model \NYY{or phenomenological model} is helpful to clarify underlying
 mechanisms of the structural formation\cite{Kennedy:2001}.
 }
 This issue is called model inadequacy and is one of the biggest
 challenges in model estimation\cite{Smith:2013}.
}

Automatic discovery of a model equation is a recent key topic in
data-driven science\cite{Daniels:2015,Brunton:2016,Brunton::2019}, and several methods have been
proposed for PDEs of time-series data \cite{Voss:2004,MUeLLER:2004,Rudy:2017,Zhao:2020}.
\NY{
 These approaches are designed to handle time-series data and cannot be applied to estimation for a stationary
pattern, because it does not have information about trajectories from
their initial
conditions.
In addition, the previous studies focus on the estimation for the data
with ground truth.
The data is generated typically from numerical results of a known model
with known parameters, and then, those parameters are estimated from
the data.
\NYY{
Therefore, the two issues, the estimation for stationary data and data
without ground truth, remain elusive even in a technical aspect of model discovery.
}
}

\NY{To overcome these issues, in this study}, an inverse problem is formulated to reproduce a given
 snapshot of a pattern, after which we propose a method,
 referred to as Bayesian modelling of partial differential
 equations (BM-PDE), to identify the best \NYYY{dynamical} PDE model and its parameters.
 \NY{
We apply our method to the problem without ground truth.
 }
We prepare the target pattern according to crystallographic
symmetry, which is independent of any candidate models, and then
perform an estimation of the best model.
\NY{
We demonstrate the BM-PDE for complex patterns such as QC, DG, and FK A15
patterns.
}
\NYY{
We also demonstrate that from the estimated parameters,
three-dimensional dodecagonal QCs can be generated.
\NYY{
The success shows a potential application of BM-PDE to understand a
mechanism of structural formation of novel materials.
 }
}

 \section*{Basic Formula}

\begin{figure}[H]
 \begin{center}
  \includegraphics[width=0.85\textwidth]{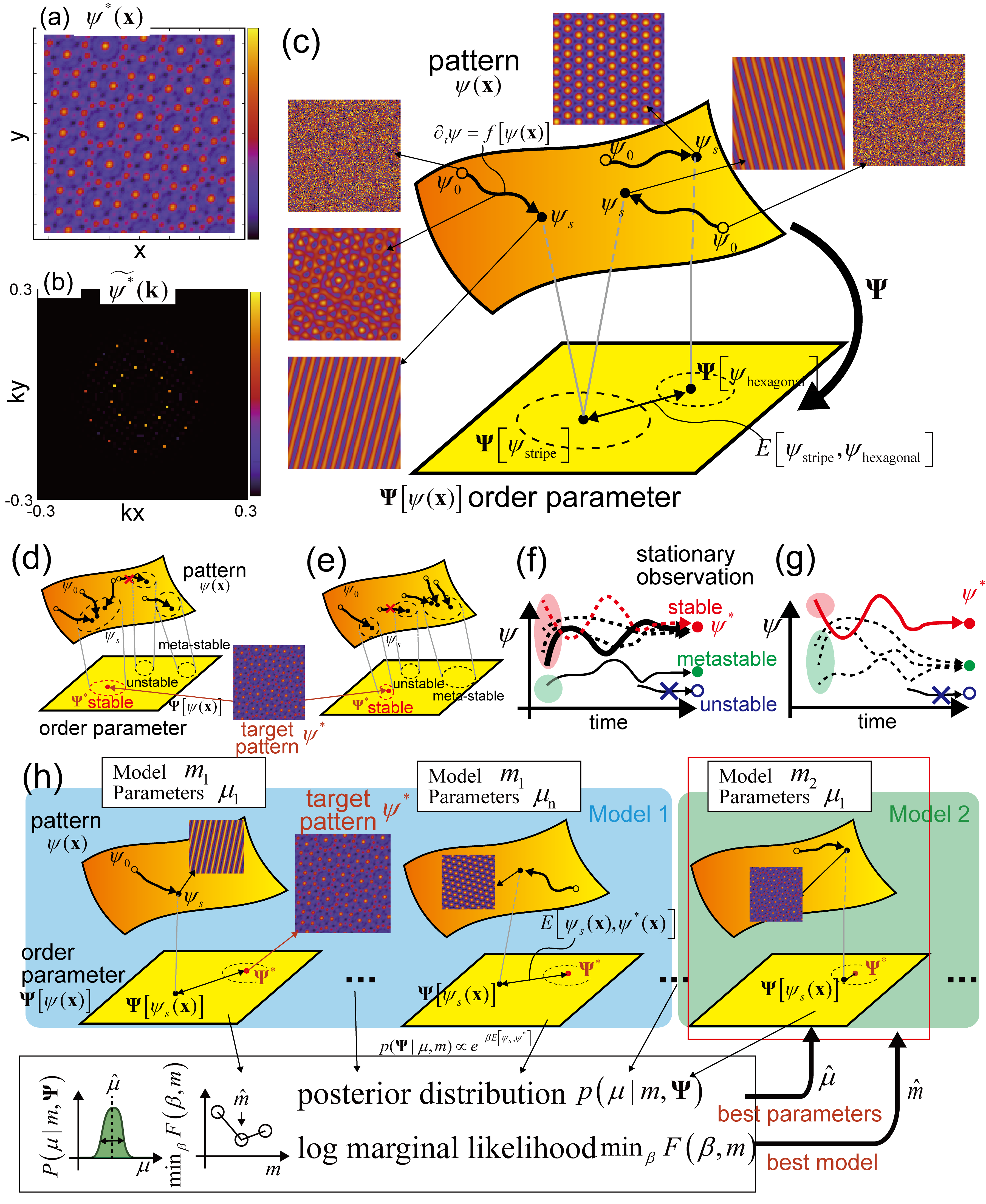}
  \caption{
	     Schematic illustration of Bayesian modelling
  of partial differential equations (BM-PDE).
	     (a,b) The example of the target pattern of a dodecagonal QC produced from a numerical result (a), and its
	     Fourier transform (b).
	     The colour bar indicates $[-1.5,3.5]$ in (a) and
	     $[0,3000]$ in (b).
  (c)  The space of patterns $\psi ({\bf x})$ and order parameters
  $\bm{\Psi}[\psi ({\bf x})]$.
  The PDE is solved with the initial condition $\psi_0$ taken from random
  variables.
  For each trajectory, the translational
  position and orientation of the pattern $\psi_s$ would change even under exactly the same parameters and the same model.
  The order parameter identifies the two patterns by extracting
  symmetries of the pattern.
  The distance between two patterns is quantified by $E$.
  \NY{
  (d-g) Stationary observation and initial conditions.
  A pattern may be stable, meta-stable, or unstable.
  The unstable pattern cannot be generated from an initial condition.
  The better model (d,f) has broader initial conditions that generate
  stable patterns, than the worse model (e,g).
  }
  \NY{(h)} For each model and each set of parameters, there is a
	     stationary pattern $\psi_s$.
	     The cost function $E[\psi^*,\psi_s]$ (energy) is calculated from the
  order parameters of the target and generated patterns.
  From the posterior distribution, the best parameters and their errors are estimated. 
	     The distribution of the cost function gives the log
  marginal likelihood of the model from which
  the selection of models can be made.
\label{fig_schematic}
}
 \end{center}
\end{figure}

We consider a pattern (or crystalline structures) expressed by the
scalar density field $\psi({\bf x})$.
An example of a \NYY{two-dimensional} dodecagonal QC \NYY{(DDQC)} is shown in Fig.~\ref{fig_schematic}(a,b).
Higher density spots may be considered as a position of particles.
\NYY{
We estimate a \NYYY{dynamical} model to reproduce a target pattern $\psi^*({\bf
x})$ as a stable pattern at
the steady state $\psi_s({\bf x})$ of a nonlinear partial differential
equation $\partial_t \psi ({\bf x}) = f_{\mu}[\psi ({\bf x})]$ (Fig.~\ref{fig_schematic}(c-e)).
If the PDE and its parameters $\mu$ are ground truth for the target pattern,
$\partial_t \psi^*({\bf x})=f_{\mu}[\psi^*({\bf x})]=0$ is satisfied.
}
\NYY{
Our target pattern is {\it one} snapshot and has information only about
the stationary state (Fig.~\ref{fig_schematic}(d-g)).
Its transient pattern from the initial state to the stationary state is
not available.
We assume the stationary state is stable in the sense that the pattern
is generated from a broad range of the initial conditions (Fig.~\ref{fig_schematic}(f,g)).
This assumption is natural when the target pattern is obtained from an experimental
result; the pattern should be reproducible.
}

\NYY{
At a first look, \NYY{the estimation for the stationary data is} impossible.
When $f_{\mu}[\psi^* ({\bf x})]=0$ is a true model, we
may have a series of {\it equally true} models such as $f_{\mu}[\psi^* ({\bf
x})]^2=0$, and $\left( f_{\mu}[\psi^*
({\bf x})] + 1 \right)f_{\mu}[\psi^* ({\bf x})]=0$.
Therefore, the estimation is not unique.
Nevertheless, as we will see later, the estimation of a model that reproduces the
target pattern as a stable structure plays a role as regularisation
 (see
 also Supplementary Sec.~\ref{SM.sec.relatedworks} \cite{seeSM} for the comparison
 with other approaches)
 }.
\NYY{
We should note that our problem is not the parameter estimation for
$f_{\mu}[\psi^*({\bf x})]=0$ in which the stability of the stationary state is
not guaranteed.
}

\NYY{
To see the difficulty of estimating stationary data, it is
instructive to consider the state-space model, widely used in data assimilation\cite{Evensen:2009,Law:2015}.
 The cost function $E$ consists of measurement (observation) and model
 errors, and is expressed as
 \NYY{
 \begin{align}
  E [\mu, \psi(\bf x)]
  &=
  \frac{1}{2}
  \|
\psi^* ({\bf x}) - \psi ({\bf x})
  \|
  +
  \frac{1}{2}
  \|
\partial_t \psi ({\bf x}) - f_{\mu}(\psi({\bf x}))
  \|
  \label{state.space}
  .
 \end{align}
 }
 In many cases, the norm $\|\cdot\|$ is chosen to be the square norm.
 If the observation contains an error, we have to estimate both the
 parameters $\mu$ and the state $\psi({\bf x})$.
 When the model represents a deterministic system, the state is described by its initial
 condition $\psi_0({\bf x})$, and accordingly the cost function becomes
 $E[\mu, \psi_0({\bf x})] $.
 When the observation does not contain noise, the first term in equation (\ref{state.space})
 vanishes, and the problem falls into a simple regression (see
 also Supplementary Sec.~\ref{SM.sec.relatedworks} \cite{seeSM}).
In the conventional data assimilation, both $\mu$ and $\psi_0$ are
 estimated by minimising $E[\mu, \psi_0({\bf x})] $ \cite{Law:2015}.
 However, in the problem of pattern formation, the specific
 initial condition to produce the pattern is not of our interest.
Moreover, there are many initial conditions that asymptotically reproduce the same
 pattern (Fig.~\ref{fig_schematic}(f)).
 Therefore, the estimation under a given stationary target pattern cannot be unique.
 In the case of time-series data, the estimation is possible because
 each trajectory is from a different initial condition.
 In our method, we sample the stationary pattern by solving the model
 under each parameter and each realisation from the random initial
 condition.
 By marginalising the initial conditions, we may obtain a unique estimation.
}

 The basic structure of our estimation is schematically shown in \NYY{Fig.~\ref{fig_schematic}(h)}.
 \NYY{Parameters are estimated from the posterior distribution under the
 target pattern, whereas the best model is estimated from the log
 marginal likelihood. }
 For a given model $m_i$ and parameters $\mu$ in a PDE, the stationary
pattern is uniquely determined under each initial condition, $\psi_0
({\bf x})$ (Fig.~\ref{fig_schematic}(c)).
We treat an initial pattern as a latent variable that is marginalised
using a random variable for the initial condition.
This is because the obtained pattern may be translationally shifted or
rotated by changing an initial pattern (see Fig~\ref{fig_schematic}(c)).
We quantify the similarity between two patterns $\psi_1$ and
$\psi_2$ by the distance between them defined as
\begin{align}
 E [\psi_1,\psi_2]
 &=
 \left|
 {\bf \Psi} [\psi_1]
 -
 {\bf \Psi} [\psi_2]
 \right|^2
 .
 \label{cost.function}
\end{align}
Our target pattern is ordered and has many invariants; the pattern must
be identified under change by translation and rotation, and also the
pattern does not change by the action of the symmetry group that the
pattern has \NY{(see Fig.~\ref{fig_schematic}(a,b))}.
Here we introduce {\it order parameter} \NYY{${\bf \Psi} [\psi({\bf
x})]= \left( \Psi_1[\psi({\bf x})], \ldots \Psi_{l_0}[\psi({\bf x})] \right)$}, which
maps the pattern onto 
the feature space and eliminates the redundant information of the ordered pattern $\psi({\bf x})$ due
to symmetry (see Fig.~\ref{fig_schematic}(c) and Methods).
The distance defined by equation (\ref{cost.function}) identify the patterns up to
symmetry transformation thanks to the order parameter.

\subsection*{Bayesian modelling}

Our goal is to find the most
 probable model $\hat{m}$ \NY{described by a PDE and its parameters
 $\hat{\mu}$ for a given target pattern $\psi^*({\bf x})$.
 We also want to quantify the uncertainty of the estimation}.
 To achieve this, we use the cost function $E [\psi_s,\psi^*]$, or called the energy, expressed by the order
 parameter \NYY{${\bf \Psi}$}, and compute the distance from the target pattern, ${\bf \Psi}^* = {\bf
 \Psi} [\psi^*({\bf x})]$, to the numerically generated
 stationary pattern for each model and parameter set, ${\bf \Psi}[\psi_s
 ({\bf x})]$.
 Our purpose is not to estimate specific initial states $\psi_0$ for the target pattern $\psi^*$, but to estimate the best
model that could generate patterns similar to $\psi^*$ independent of the initial state.
Therefore, our best parameter set $\hat{\mu}$ is defined by the mean of the marginal probability
distribution \NYY{under a model $m$}
 \begin{align}
  p(\mu \mid {\bf \Psi}^*, \beta, m) = \int p(\psi_0) p(\mu \mid {\bf
  \Psi}^*, \psi_0, \beta, m) d\psi_0.
  \label{marginalisation.IC}
 \end{align}
 \NY{
The integral over the initial conditions $\psi_0$ implies that the
posterior distribution of the parameters is chosen so that the estimated
parameters can generate the target pattern from a wide range of the
initial conditions.
We may avoid the parameters that can generate the target pattern only for a specific
initial condition (Fig.~\ref{fig_schematic}(f,g)).  
 }
Following Bayes' theorem, the posterior distribution \NYY{under a fixed $\psi_0$} is given by
 \begin{align}
  p(\mu \mid {\bf \Psi}^*, \psi_0, \beta, m)
  &=
  \frac{p({\bf \Psi}^* \mid \psi_0, \mu, \beta, m) p(\mu \mid m)}{p({\bf \Psi}^* \mid m, \beta)}. 
 \end{align}
The likelihood is represented by $p( {\bf \Psi}^* \mid \mu, m, \beta)
 \propto e^{-\beta E[\psi, \psi^*]}$, where the inverse temperature
 $\beta$ is associated with variance of the observation noise.
This likelihood implies that the error in the measurement is given by ${\bf \Psi}^* = {\bf \Psi} + \bm{\xi}$ with the Gaussian
 noise $\bm{\xi}$ with zero mean and its variance $\beta^{-1}$.
 The prior distributions $p(\psi_0)$ and $p(\mu \mid m)$ are
 \NYY{assumed to} the uniform distribution.
 The normalisation factor $p({\bf \Psi}^* \mid \beta, m)$, or the log
 marginal likelihood (free energy) $F(\beta, m) \equiv -\log p({\bf
 \Psi}^* \mid \beta, m)$, is one of the criteria of model selection and
 hyperparameter estimation \cite{mackay1992bayesian,Bishop:2006}.
 \NY{
 Both $p({\bf \Psi}^* \mid \beta, m)$ and $F(\beta,
 m)$ play a role as a probability density of the models and
 hyperparameters (see Methods).
Therefore, our best model $m$ and inverse temperature $\beta$ are both determined by
 maximising $p({\bf \Psi}^* \mid \beta, m)$, or equivalently, minimising
 $F(\beta, m)$ \cite{Tokuda:2017}.
}


\subsection*{\NYY{target patterns without ground truth}  }
\NY{
In this work, we consider two types of target patterns; (i) a numerical
solution of the PDE model that we use, and (ii) a pattern synthesised by
superposition of plane waves as equation (\ref{phi.cos}).
The former has ground truth, whereas the latter does not.
The synthesised pattern is independent of our PDE models, and therefore,
it is not necessarily a solution for the PDEs.
 We will demonstrate that our approach still works for the problem
 without ground truth.
 This problem is an intermediate step between the estimation of the
 problem with ground truth and experimental data.
 We do not know the true PDE for experimental data.
 Still, we would like to estimate the best PDE model to explain the data to
 understand the mechanism of the structural pattern formation.
\NYY{
 Our approach has the advantage in that we do not know {\it a priori} the model that
 reproduces the target pattern.
 In most of model identification, the problem with ground truth has
 been used.
 }
 }

\begin{figure}[H]
 \begin{center}
  \includegraphics[width=0.90\textwidth]{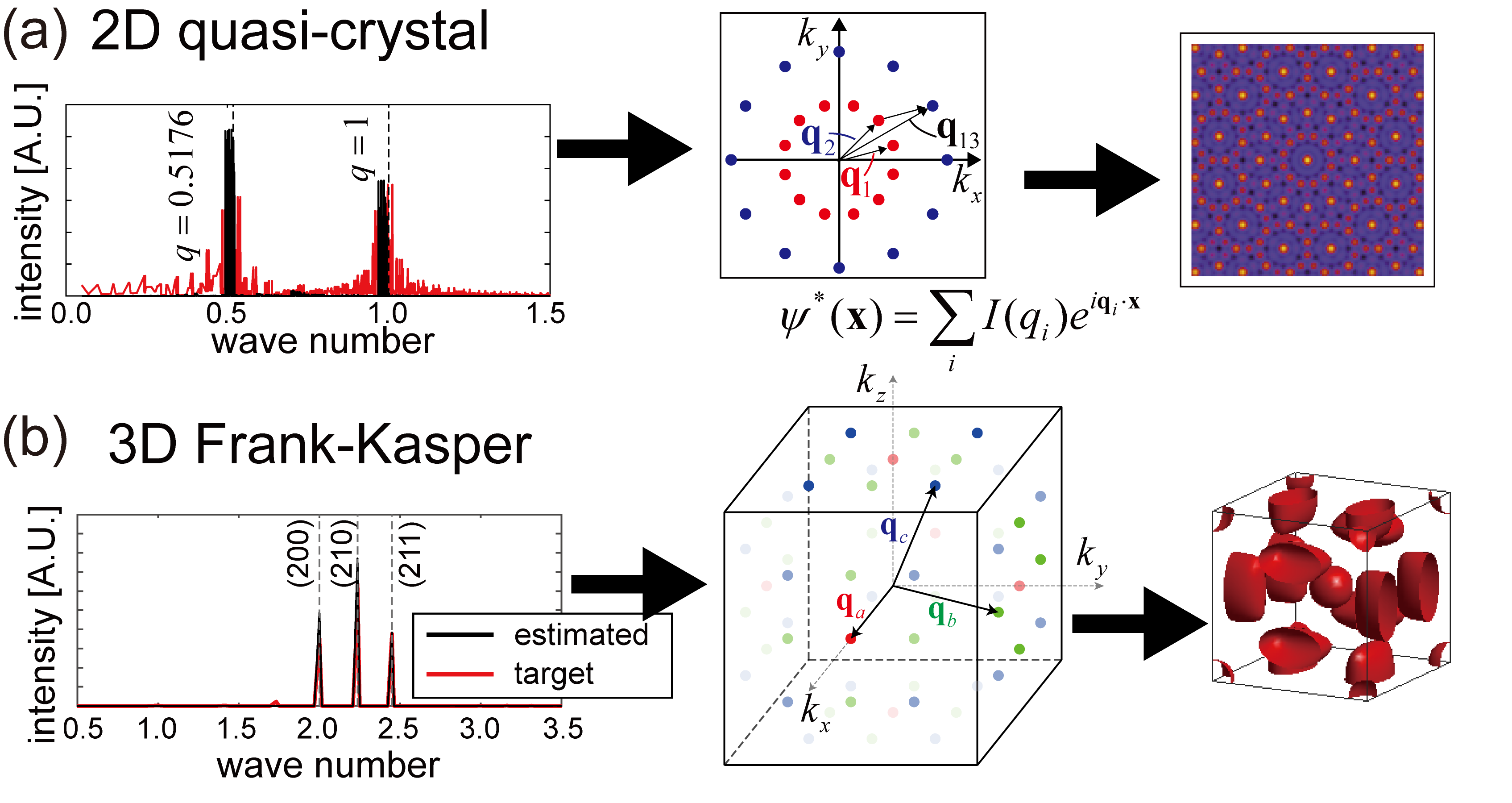}
  \caption{
  \NY{
  The way to synthesise the target patterns without ground truth by equation (\ref{phi.cos}). (a)
  two-dimensional dodecagonal QC and (b) FK A15.
  (left) The diffraction patterns of the target and estimated patterns.
  (middle) From the peaks of the diffraction pattern, the spots in the
  Fourier space are identified for each wave mode.
  In (a), the spots correspond to ${\bf q}_i = q_0 \left( \cos
  (2i+1)\pi/12, \sin (2i+1)\pi/12 \right)$ for $i=1 \ldots 12$, and ${\bf q}_i = q_1 \left( \cos
  i\pi/6, \sin i\pi/6 \right)$ for $i=13 \ldots 24$.
  The ratio between the two wavelength
  $q_0/q_1=1/(2\cos(\pi/12))=0.51\cdots$.
  In (b), the spots indicate ${\bf q}_a = (2,0,0), (0,2,0), \ldots$ in
  red points, ${\bf q}_b = (2,1,0),(1,-2,0), \ldots $ in green points,
  and ${\bf q}_b = (2,1,1),(1,1,-2), \ldots $ in blue points.
  (right) By superposition of the waves in the Fourier space, the real space target
  patterns are synthesised by $\psi ({\bf x}) = \sum_i I(q_i) e^{i {\bf
  q}_i \cdot {\bf x}}$, where $I(q_i)$ is the intensity of the wave of
  ${\bf q}_i$, and the sum is taken over all the spots.
  }
  \label{fig_notruth}
  }
 \end{center}
\end{figure}

 \NY{
 To synthesise the target structure by equation (\ref{phi.cos}), positions of the peaks and their
 amplitude in the Fourier space are required (Fig.\ref{fig_notruth}(a)).
 These quantities can be directly measured in the scattering experiments
 in two dimensions and three dimensions.
 The positions and the amplitude in
 the Fourier space can be reproduced from the structure factor, as shown
 in Fig.\ref{fig_notruth}.
 We use this information to synthesise the target.
 \NYY{
 We demonstrate the estimation of PDEs for the Frank-Kasper A15
 structure taken from the experimental data\cite{Jayaraman:2018}
 (Fig.\ref{fig_notruth}(b)).
 The detailed method to synthesise the target structure is shown in
 Methods.
 }
 }

 \section*{The estimation for phase field crystal models}

 \NYY{
 We demonstrate our method for a class of phase-field crystal models.
 Specifically, we make a family of the models by changing mainly the linear
 part of equation (\ref{eq.PDE}).
 Our approach is not limited to the choice of this family.
 In fact, we can replace equation (\ref{eq.PDE}) with any other families of models.
 A possible extension of this approach, including the estimation of
 nonlinear terms, is discussed in Supplementary
 Sec.\ref{sec.family.model} \cite{seeSM}.
 }
 We consider a model, called $m \in M$, expressed by \NYY{a nonlinear PDE} of the form
\begin{align}
 \partial_t \psi ({\bf x})
 &=
 \mathcal{L}_{\mu}^{(m)} \psi ({\bf x})
 + \mathcal{N} \left[ \psi ({\bf x}) \right]
 \label{eq.PDE}
\end{align}
with a set of parameters $\mu$.
The PDE is decomposed into two parts.
The linear term is expressed by the linear operator,
$\mathcal{L}_{\mu}^{(m)}$, acting on $\psi({\bf x})$.
 \NY{
Because we are interested in patterns in bulk, not affected by
boundaries, we use periodic boundary conditions.
 }

We make a family of models, \NY{$M=\{ m_i \}_{i = 1,2,\ldots  ,i_{\rm
max}}$}.
\NYY{
In model $m_i$
the linear operator has $i$ peaks in its spectrum (see
Fig.~\ref{fig2D}(f-h)).
}
Our family of models is designed to have a physical interpretation that the system has \NYY{$i$
length scales for model $m_i$} because peaks in the spectrum correspond to the number of length scales.
Each length scale is characterised by its wavelength $q_i$ and
the value of its spectrum at the wavelength $a_i$ (\NY{see
 Fig.~\ref{fig2D}(f-h)}).
 \NY{
 We also use the mean density $\bar{\psi}$ and system sizes in each direction as parameters.
 Note that $\bar{\psi}$ is identical to the parameter for the nonlinear
 $\psi^2$ term in equation (\ref{eq.PDE}).
 }

\subsection*{two-dimensional QC with ground truth}
To give better insight into the BM-PDE, we first focus on an example of
a two-dimensional QC with 12-fold symmetry \NYY{(DDQC)} shown in Fig.~\ref{fig_schematic}(a).
This pattern has been studied using a model with two length scales\cite{Lifshitz:1997}.
The target pattern in this section is numerically produced with a set
of parameters $\mu^*$.
The two-length-scale model is used, that is, $m^*=m_2$.

\begin{figure}[H]
 \begin{center}
  \includegraphics[width=1.00\textwidth]{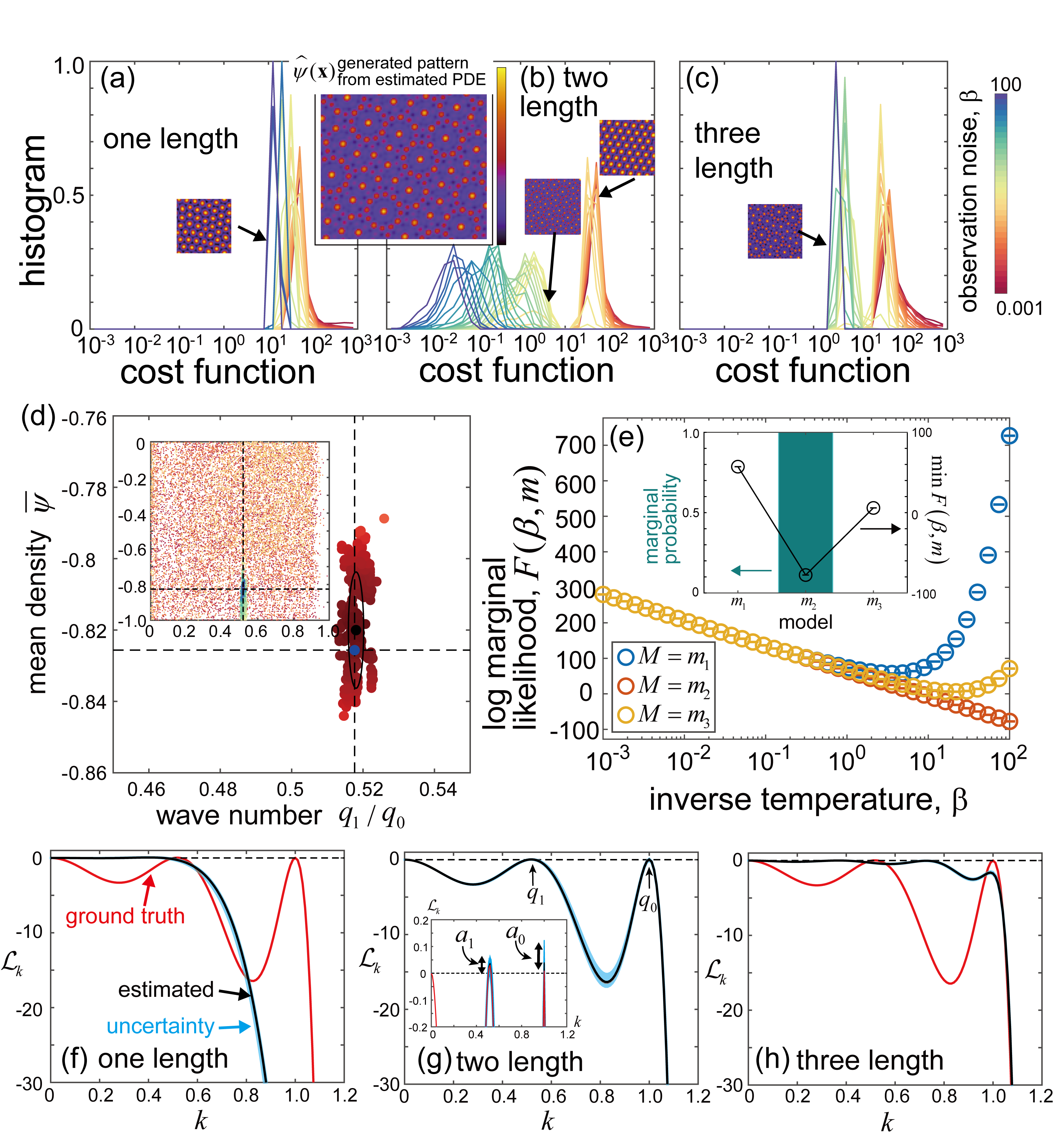}
  \caption{
  Model selection and parameter estimation for the target pattern of
	     two-dimensional QC pattern with 12-fold (dodecagonal) symmetry. 
	     (a-c) The histograms of the cost function, $E[\psi^*,\psi_s]$, during the
  sampling.
  The horizontal axis is shown in the logarithmic scale.
    The generated pattern from the estimated PDE is shown in the insets.
  Typical patterns at each energy range are also shown in the insets with
  arrows.
	     (d) The estimated parameters in the space spanned by $q_1$ and $\bar{\psi}$.
	     The colour indicates \NY{a histogram} where darker red corresponds
	     to \NY{a higher} probability.
	     The mean and standard deviation of the estimated parameters
	     are shown in the black point and the black line,
	     respectively.
	     The ground truth parameter values are shown in dashed lines and the
	     blue point.
	     The inset shows the same plot under various $\beta$ in the range of parameters
  used for the prior distribution.
  The same colour code as (a-c) is used. 
	     (e) Model selection is made by the log marginal
  likelihood (free energy)
	     calculated from the steady state energy distribution for
  $m_1$ (a),  $m_2$ (b), and $m_3$ (c).
  The inset shows the probability of each model marginalised for all
  $\beta$.
  The minimal free energy of each model is also shown with error bars,
  which overlap with the points.
  \NY{
  (f-h) The linear spectrum (black lines) as a function of wavenumber from the
  estimated parameters for $m_1$ (f),
  $m_2$ (g), and $m_3$ (h).
  The ground truth is shown in red lines.
  The uncertainty of the estimation is shown by the range in light blue.
  Note that in (g) the estimated line and ground truth are overlapped.
  The inset in (g) shows the same plot near
  the peaks.
  }
\label{fig2D}
}
 \end{center}
\end{figure}

 For $m=m_2$, the \NY{cost function} $E[ \psi_s, \psi^*]$ decreases
 \NYY{during the sampling from the posterior distribution}, and the generated stationary patterns from equation (\ref{eq.PDE}) converge to QCs which
 are similar to the target pattern (Fig.\ref{fig2D}(b), see also
 Supplementary Fig.\ref{SM_fig2D} \cite{seeSM}).
The estimated parameters well agree with the parameters
that we used to generate the target pattern (see Fig.\ref{fig2D}(d) and
 Supplementary Table.\ref{estimated.parameters.numerical.12fold} \cite{seeSM}).
The \NY{estimated} length scale is $q_1=0.52$ with the second length scale
$q_0=1$.
The ratio between them agrees to $q_0/q_1 = 2
 \cos(\pi/12) \approx 1.9319$, which is the known
value to generate this pattern\cite{Lifshitz:1997}.
The BM-PDE automatically estimates this ratio starting from \NYY{uniform} prior distribution of the wavelength.
The estimation also works for other parameters
(Supplementary Table.\ref{estimated.parameters.numerical.12fold} and
 Supplementary Fig.\ref{SM_fig2D} \cite{seeSM}).

Using the estimated parameters, we may generate a pattern similar to the
 target pattern from uniform
 random initial density (see the inset of Fig.\ref{fig2D}(b)).
To see the quality of the estimation, we measure the steady distribution
of the \NY{cost function}.
Figure ~\ref{fig2D}(b) shows that
there are two distinct states:
one has \NY{a higher cost function} $E \gtrsim 10^2$ and the other has
 \NY{a lower cost function} $E
\lesssim 10^2$.
The latter corresponds to QC, whereas hexagonal patterns
mainly dominate the former.
The gap between the two states indicates that the QC patterns
require a high resolution in the parameter search.
The large step in the parameter space cannot find the optimal
parameters because their range is narrow in the prior range of the parameters as in the inset of Fig.\ref{fig2D}(d).
Therefore, the conventional gradient method with a fixed step
size either cannot find the QC when the step size is large,
or impractical when it is small.   
The BM-PDE use hierarchical sampling, such as replica exchange Monte
Carlo (REMC) \cite{Geyer:1991,Hukushima:1996}.
In this method, the knowledge of the step size corresponding to the
\NY{observation noise} $\beta$ is not a prerequisite for the parameter search. 

The same algorithm is performed for the models with one length scale
$m=m_1$ and three length scale $m=m_3$.
It is not surprising that the cost function $E$ of the one-length-scale
model is much higher
than that of $m_2$ because it cannot reproduce a
QC pattern (Fig.~\ref{fig2D}(a)).
In fact, the model $m=m_1$ with the estimated parameters produces
hexagonal patterns rather than QCs.
The three-length-scale model does reproduce QCs,
which are comparable to the target pattern (Fig.~\ref{fig2D}(c)).
However, our Bayesian estimation selects the two-length-scale model.
Figure \ref{fig2D}(e) demonstrates that the minimum free energy, that is,
\NY{the marginal likelihood, is lower for $m=m_2$}.
Therefore, we estimate that this QC pattern is described by the interaction
of two modes with different length scale.

\subsection*{\NYY{various target patterns} without ground truth}

\begin{figure}[H]
 \begin{center}
  \includegraphics[width=1.00\textwidth]{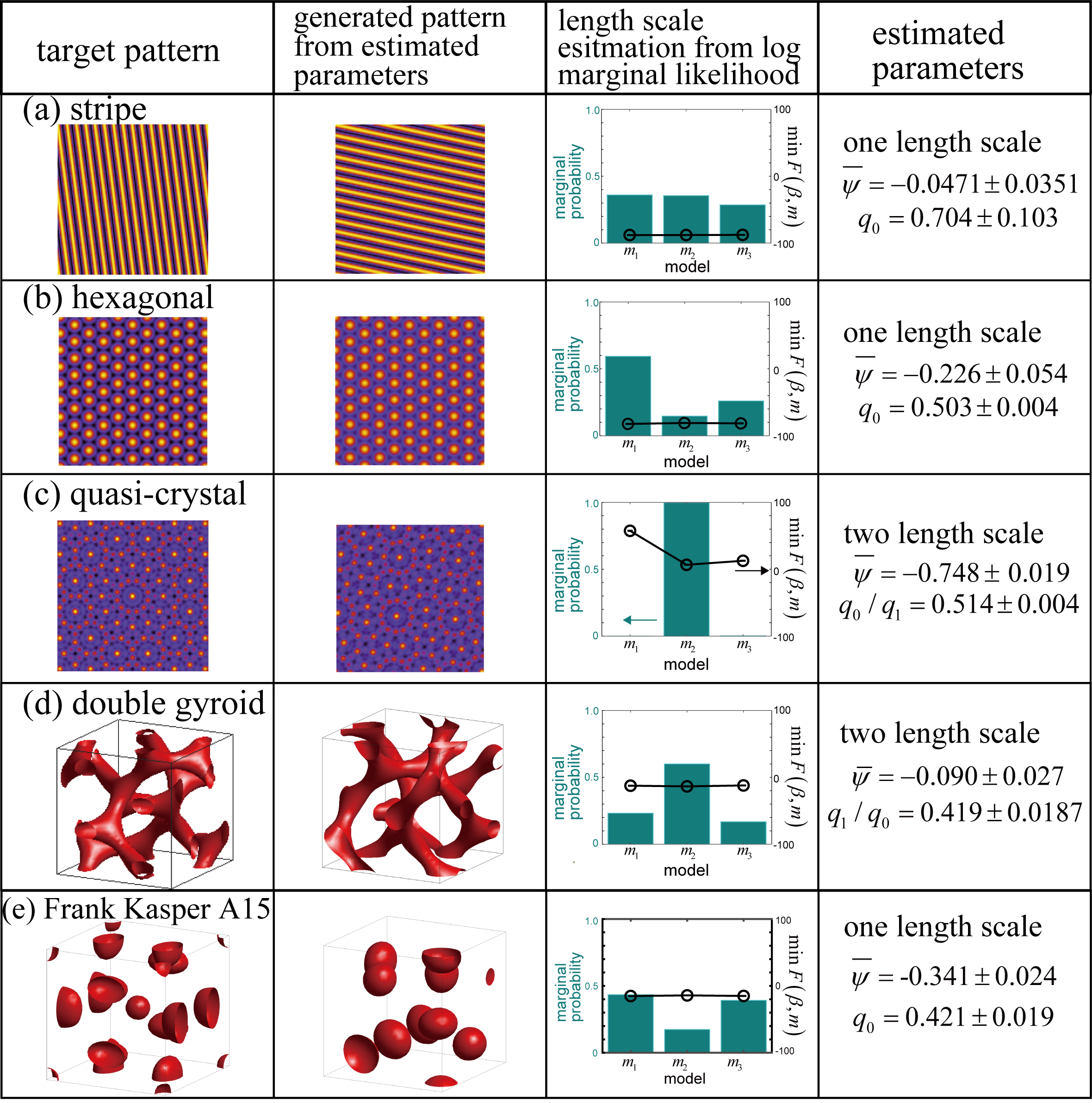}
  \caption{
	     Summary of target and estimated patterns for stripe (a),
	     hexagonal (b), 12-fold symmetric QC (c),
	     \NY{DG} (d), and \NY{FK A15} (e).
	     For each pattern, the free energy is evaluated for a model
	     with one-,
  two-, and three-length scales.
  The best model is selected from minimum free energy.
 The model uncertainty is quantified by the marginal probability of each model
  obtained from the free energy marginalised for all temperatures. 
  \NY{
All the target structures are synthesised from analytical functions by equation (\ref{phi.cos});
  they do not have ground truth in the model.
 The target structure of Frank Kasper A15 is synthesised based on the
  data in experiments in \cite{Jayaraman:2018}.
  }
\label{table1}
}
 \end{center}
\end{figure}

The BM-PDE is not restricted to the estimation of the parameters that are used
to generate the target pattern.
Using a
two-dimensional \NYY{DDQC}, we demonstrate that the BM-PDE
successfully estimates the best model and approximated parameters for the
target pattern without ground truth. 
The \NYY{DDQC} is synthesised by the superposition of 12 plane waves
\NYY{in equation (\ref{phi.cos})} (see Methods).
The pattern is similar to numerically produced QC used in
the previous section (see Fig.~\ref{fig_schematic}(b) and
\NY{Figure}~\ref{table1}(c)), but in the current case, there are no ground truth
parameters and a true model.
The target pattern can only approximately be
accessed by one of the models in equation (\ref{eq.PDE}).
In contrast with the numerically produced pattern, estimated parameters
do not reproduce {\it exactly} the same pattern as the target pattern, and
therefore the \NY{cost function} is relatively high (Supplementary
Fig.~\ref{SM12fold} \cite{seeSM}).
 Nevertheless, both two-length-scale and three-length-scale models
 reproduce \NYY{DDQC} patterns.
 The estimated parameters reproduce the inherent ratio of the length scales
 $q_0/q_1= 1.948 \approx 2 \cos(\pi/12)$.
 The \NYY{marginal likelihood} indicates that the two length scale is favourable (\NY{Figure}~\ref{table1}(c)).

We summarise the results of a variety of patterns in \NY{Figure}~\ref{table1}.
For each target pattern, we can reproduce visually similar patterns, and
the most probable number of length scales.
In two-dimensional systems, stripe and hexagonal are the two
most popular patterns under one length scale. \NY{These patterns} are obtained
from the conventional
phase-field crystal model \cite{Elder:2004}.
The \NYY{marginal likelihood} calculated in the BM-PDE indeed estimates that one
length scale is favourable (\NY{Figure}~\ref{table1}(a,b)).
The difference between stripe and hexagonal patterns appears in the mean
density $\bar{\psi}$.
A quadratic nonlinear term is necessary to reproduce hexagonal patterns,
and this implies that it appears at $|\bar{\psi}| \gg 0$\cite{Elder:2004}.
 When $\bar{\psi} \simeq 0$, stripe patterns appear.
The estimated values of the mean density are consistent with the results
from \NYY{the phase diagram reported in literature}\cite{Elder:2004};
we obtain the estimated mean density $\hat{\bar{\psi}} \simeq -0.23$ and $\hat{\bar{\psi}} = -0.05$  for
the hexagonal and stripe target patterns, respectively (see also
Supplementary Fig.~\ref{SMhist.stripe} and \ref{SMhist.hexagonal} \cite{seeSM}).
The BM-PDE automatically estimates appropriate parameters from an artificially synthesised snapshot
of the target pattern.
 
\subsection*{double gyroid and Frank Kasper patterns }

We discuss an application of the BM-PDE to nontrivial three-dimensional patterns.
The target patterns are double gyroid (DG), shown in Fig.~\ref{table1}(d)
and Frank Kasper (FK) A15, shown in Fig.~\ref{table1}(e).
The DG structure has two separate domains, each of which has degree-three vertices\cite{Scherer:2013}. 
The DG structure was predicted
theoretically\cite{Podneks:1996,Shi:1999} and numerically
\cite{Zhang:2008}, but mainly discussed in similar but \NY{a different class} of
models of block copolymers\cite{nonomura:2001}.
FK A15 has been studied in metallic alloy and soft materials\cite{Bates:2019}.
For example, the self-consistent field theory, designed to describe block copolymers,
is capable of reproducing this pattern, but to our knowledge, this
structure has not been reported within the framework of PFC.

In three dimensions, the order parameter may be defined similarly to that in
two dimensions (see Methods).
Structural candidates in three dimensions are far richer than
two-dimensional patterns.
In fact, during the sampling, cylindrical patterns with a hexagonal
lattice, BCC, and other patterns appear.
 These patterns can be stable under a certain region of parameter space\cite{Jaatinen:2010}.
The BM-PDE can reproduce both the DG and FK A15 patterns \NYYY{in all
models ($m=m_1,m_2,m_3$)}.
The generated structure is the same as the target pattern, which is
evident from the peaks in the Fourier space (\NY{Figure}~\ref{table1}(d,e) and
\NYY{see also} Supplementary Sec.~\ref{SM.sec.gyroid} \cite{seeSM}).
The real space structures of the two patterns are overlapped by translation. 
The \NY{log marginal likelihood} in Fig.~\ref{table1}(d,e) shows that the target patterns
of DG and FK A15 are
expressed \NYYY{most likely} by the two length scale and one length scale, respectively.

The DG structure has two length scales with their
ratio $1.15$ (see \NY{also} Supplementary Sec.~\ref{SM.sec.gyroid} \cite{seeSM}).
In the one-length-scale model, by taking $a_0 \gg 0$, several modes with slightly different length
scales may become unstable so that the double gyroid pattern is realised.
In fact, it was discussed that DG appears not at $a_0 = \epsilon \approx 0$
with the small number $\epsilon$, but at $a_0 \gtrsim
\mathcal{O}(1)$\cite{Podneks:1996,Shi:1999}.
In addition, this pattern appears between the stripe patterns and
cylindrical (hexagonal) patterns in the phase diagram, namely $1 \gg |\bar{\psi}| \gtrsim 0$.
On the other hand, in the model of block copolymers, the FK A15 phase
appears near BCC patterns in the phase diagram \cite{Bates:2019}.
This suggests that $|\bar{\psi}| \gg 0$ to obtain FK A15.
The estimated mean density $\bar{\psi}$ well agrees with this tendency. 
(Supplementary Figs.~\ref{SMhist.gyroid} and \ref{SMFKA15} \cite{seeSM}). 
The two models $m_1$ and $m_2$ have a comparable probability for both patterns.
 The DG pattern chooses $m_2$ possibly because $m_2$ has a wider range of spectrum
 amplitude in the phase diagram.
 \NYY{
 The target structure of FK A15 has ellipsoidal domains.
 The generated structures with the estimated parameters are also
 ellipsoidal (see Supplementary Fig.\ref{SMsphericity} \cite{seeSM} for quantitative
 results on sphericity).
 Such deformation is also reported for FK A15 structure made of block copolymers\cite{Lee:2014}.
 }

\NY{
We should note that the target structure of the FK A15 is
taken from the experimental result \cite{Jayaraman:2018}.
From the X-ray scattering data, the position and
amplitude of peaks in the Fourier space are identified
(Fig.\ref{fig_notruth}(b)).
We use the positions of the peaks and their amplitude in the data in
\cite{Jayaraman:2018}.
}

\section*{Robustness against noise}
 \label{sec.robust.noise}

\begin{figure}[H]
 \begin{center}
  \includegraphics[width=1.00\textwidth]{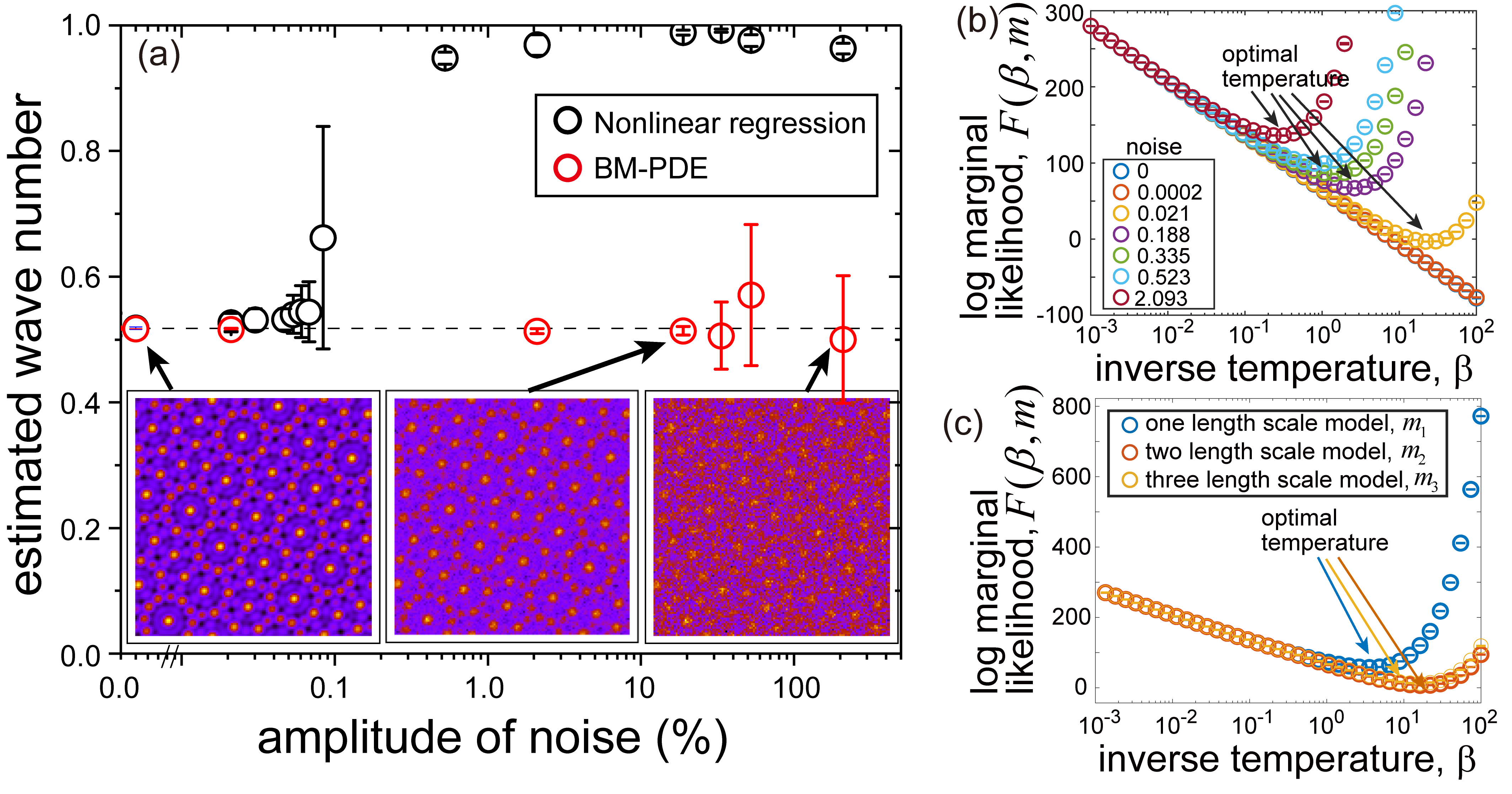}
  \caption{
  (a) Estimated wavenumbers for numerically generated QC by
  BM-PDE equation (\ref{cost.function}) (red
  points) and
  conventional regression method (black points)
  under Gaussian white noise added on the target pattern \NY{(see equation
  (\ref{nonlinear.regression}))}.
  Noise amplitude with respect to the variance of the noiseless pattern
  is defined as $\sigma^2 / {\rm Var} [ \psi^* ] $ where $\sigma^2$
  is the variance of added noise.
  \NYY{The noise amplitude is shown in \%.}
  The horizontal dashed line indicates the ground truth of the wavenumber
  $q_1=0.51764$.
  The target patterns under the different noise amplitude \NY{are shown} in the insets. 
  (b) The \NY{log marginal likelihood} at each inverse temperature $\beta$ for the target
  pattern generated by the numerical simulation with noise corresponding
  to (a). The minimums of the \NY{log marginal likelihood} are shown by arrows.
  (c) The free energy at each inverse temperature $\beta$ in REMC under
  the models $m_1$, $m_2$, and $m_3$ for the target pattern synthesised
  by the function \NYY{of equation (\ref{phi.cos})}.
\label{fig_noise}
}
 \end{center}
\end{figure}

We hypothesise that the robust estimation of the model and parameters
for the target pattern without ground truth originates from robustness
against noise. 
To see this, we add Gaussian white noise \NYY{to the target pattern in Fig.~\ref{fig_schematic}(a)} and study
its effect on the estimation of
the parameters.
\NYY{
We focus on the two-length-scale model $m=m_2$ and the estimation of
wavelength because this parameter has a narrow acceptable range.
  For the \NYY{DDQC}, the wavenumber is required to be
 close to
 $0.51$, and otherwise, the pattern cannot be generated because the mode
 coupling between two length scales does not occur.
}
Figure~\ref{fig_noise}(a) shows that even when the amplitude of the
 noise is about $20 \%$ of \NYY{the amplitude of the density field of} the target pattern, the
 estimation of the parameter successfully works.
 Beyond the noise amplitude,
 the error of the estimated wavenumber becomes large, and the fraction
 of the patterns different from the target pattern increases.

  We compare the BM-PDE with the standard regression methods in
 which parameters $\mu$ are estimated by minimising the cost function $\|
 \partial_t  \psi - f(\psi;\mu)\|$ under an appropriate norm \NYY{(see Methods)}.
 This method relies on the state noise added in the dynamical equation, and
 thus, does not give a good estimate for the measurement
 noise\cite{Voss:2004} (see \NYY{also Supplementary
 Sec.\ref{SM.sec.relatedworks}} \cite{seeSM}).
 In fact, Fig.~\ref{fig_noise}(a) shows that the estimated wavenumber deviates from ground truth even
 under $0.1 \%$ noise amplitude. 

In contrast with the noiseless target pattern, which has its \NY{log
marginal likelihood $F[\beta]$}
monotonically decreasing as \NY{$\beta$ increases} in
Fig.~\ref{fig2D}(e), the \NY{log marginal likelihood} for the target pattern with noise
has a minimum at the finite \NY{$\beta$} as in Fig.~\ref{fig_noise}(b).
The minimum demonstrates the optimal \NY{noise} corresponding to the
\NY{noise amplitude in the target pattern}\cite{Tokuda:2017}.
 The \NYY{noise level} \NY{$\beta^{-1}$} at the \NY{minimum of log marginal likelihood} increases as the amplitude
 of noise \NY{in the target} increases.
 At the large noise amplitude $\gtrsim 30 \%$, the minimum \NY{$\beta$}
 reaches a comparable value with the \NY{cost function} at the gap between two structures in
 Fig.~\ref{fig2D}(b): QC and hexagonal patterns.
 Thus, \NYY{DDQC} can no longer
 be reproduced.
 Interestingly, the minimum \NY{of the log marginal likelihood} is also observed for the target
 pattern synthesised by the function \NYY{of equation (\ref{phi.cos}) (Fig.~\ref{fig2D}(c))}.
 In this case, there is no ground truth parameter that reproduces exactly the
 same pattern as the target pattern.
 The optimal \NY{$\beta$} at the minimum \NY{of $F[\beta]$} describes the
 deviation of the target pattern from the patterns that the models can
 reach.
 Without the Bayesian inference, we cannot successfully estimate the
 uncertainty, which plays a similar role to noise.

 \NYYY{
 The BM-PDE works for the damaged target structure in which some
spatial information is lost.
We demonstrate the estimation for the damaged data in Supplementary
 Sec.\ref{SM.sec.noise} \cite{seeSM}.
The BM-PDE does not rely on the time and spatial derivatives of the
target structure.
Therefore, the damage in the data can be randomly distributed, namely the
 information of neighbouring data point is not necessary.
This is not the case if the information of spatial derivatives is
 necessary for the estimation. 
}

 \section*{Three-dimensional dodecagonal pattern expressed
 by a PDE}

\begin{figure}[H]
 \begin{center}
  \includegraphics[width=1.00\textwidth]{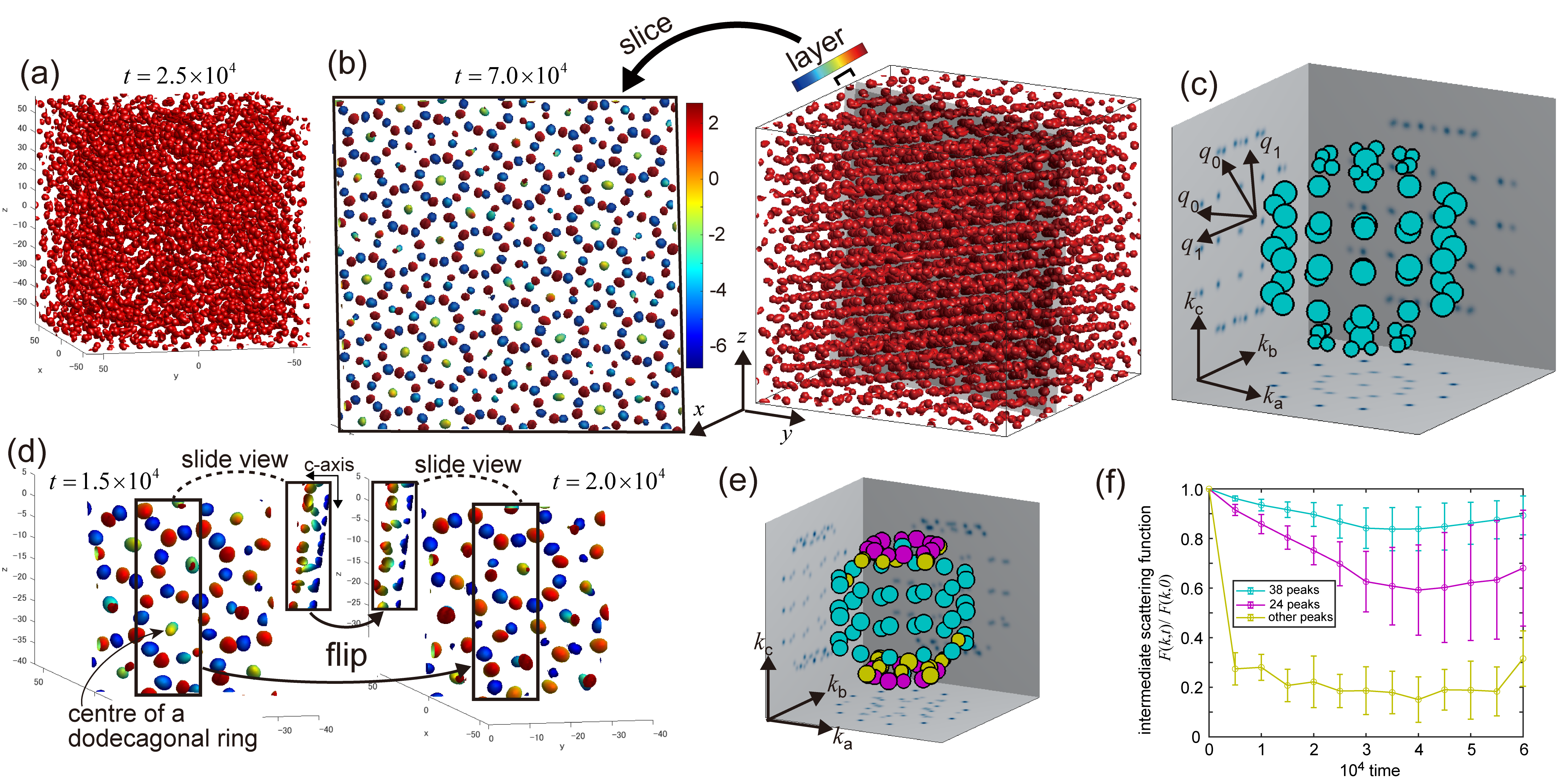}
  \caption{
  \NYY{
  Structures of the DDQC and their fluctuations.
  (a) Meta-stable micelle-like structure.
  (b) The DDQC (right) and its slice perpendicular to the axis of
  12-fold symmetry in the region of the box in black.
  The colour indicates the position along the axis.
  Therefore, the red and blue domains in the dodecagonal ring are
  located in different layers.
  The yellow domains are located between the two layers.
  (c) The DDQC in the Fourier space.
  The point sizes are proportional to the amplitude of the peak of
  $| \tilde{\psi}_{\bf k} |$.
  The structure has 12-fold symmetry around the axis of $k_c$, and $k_a$
  and $k_b$ are perpendicular axes to $k_c$.
  The projection of $| \tilde{\psi}_{\bf k} |$ on each plane is also
  shown.
  Two lengths of wave vectors $q_0$ and $q_1$ are shown.
  (d) A local structural change in the DDQC.
  The slice perpendicular to the symmetry axis is shown at two different
  times.
  The insets show an enlarged structure along the symmetry axis.
  (e,f) Fluctuations of the DDQC.
  In (e), the peaks in the Fourier space are divided into three groups
  by their amplitude.
  The 38 cyan peaks have the largest amplitude, and the 24 magenta peaks
  have the next largest. The other peaks are shown in yellow.
  The 38 cyan peaks and the 24 magenta peaks form the main peaks in (c).
 The normalised intermediate scattering function is shown for each
  group of peaks in (f).
  Each colour corresponds to $F(k,t)$ of the peaks in the same colour.
  The points are mean values in each group, and the error bars are
  standard deviation in each group. 
  }
  \label{fig_3Ddodecagonal}
}
 \end{center}
\end{figure}

 \NYY{
Using the estimated parameters, we may investigate additional physical
insights of the model.
 To demonstrate it, we consider DDQC
 in three dimensions (Fig.~\ref{fig_3Ddodecagonal}(b,c)).
 Although the icosahedral QCs have been found in the PDE model using two-length-scale PFC\cite{Subramanian:2016}, other
 three-dimensional structures are largely unexplored
 \cite{Damasceno:2017}.
 The axisymmetric dodecagonal structure has twelve-fold symmetry along
 the axis ($c$-axis) of axisymmetry (Fig.~\ref{fig_3Ddodecagonal}(b,c)).

 To generate this structure, larger system size is required, but it is
 computationally demanding.
Moreover, as we will discuss later, we found that the kinetics of the formation of the DDQC in
three dimensions is
fundamentally different from other structures.
Because the FK A15 is known as the simplest approximant of the
DDQC\cite{Iacovella:2011}, it is fair to assume that the DDQC may appear
 near the estimated parameters of the Frank-Kasper A15.
 Therefore, we focus on the estimated parameters,
  and solve the estimated models with larger systems
size in a longer time scale.

The structure discussed in the previous sections, including the DDQC in
two dimensions, may appear from a random initial condition by quenching.
The random initial structure reorganises to the desired
pattern without noise.
In addition, the system reaches its stationary state before $t=10^2-10^3$.
Note that the time scale $t=1$ corresponds to the relaxation time of the
structure when $a_i=1$.
The formation of the DDQC in three dimensions is entirely different from the
two-dimensional structure.
It requires not only a larger system
size but also an annealing process.
Without the state noise, the initial random structure forms micelle-like isotropic
structures (Fig.\ref{fig_3Ddodecagonal}(a)), which is frozen.
This is particularly the case for the system size $N>64$.
To overcome the trap at the metastable structure, we add the state noise into the
PDE model (see Methods), and anneal the structure by decreasing the noise amplitude.
We found that the annealing process is very slow; the DDQC appear around
$t\gtrsim 5\times 10^4$.

 Figure ~\ref{fig_3Ddodecagonal}(b) shows the generated structure for the
 system size $N=256$.
 The generated structure is periodic along the $c$-axis.
 We denote the other two axes as the $a$-axis and the $b$-axis, and in the
 Fourier space, these axes are denoted by $k_a$, $k_b$, and $k_c$.
 As in Fig.~\ref{fig_3Ddodecagonal}(c), the structure has twelve-fold symmetry in the plane perpendicular to
 the $c$-axis.
 The twelve-fold symmetry arises from the layered structure in which two
 layers with the hexagonal symmetry rotate $\pi/6$ with each other
 (Fig.~\ref{fig_3Ddodecagonal}(b)).
 The centre of the dodecagonal ring is located between the two layers.
  There are 62 main peaks in the Fourier space.
 Along the symmetry axis, there are five 12-fold rings and two spots
 along the $c$-axis.
 From the peaks in the Fourier space shown in
 Fig.~\ref{fig_3Ddodecagonal}(c), it is evident that there are two length scales $q_0$ and $q_1$ whose ratio is $q_1/q_0=2/\sqrt{5}$.
 This value is, in fact, close to the estimated ratio of parameters $\hat{q}_1/\hat{q}_0=0.889$. 
 The layered structure with the dodecagonal symmetry has been reported
 in particle based simulations\cite{Dzugutov:1993,Damasceno:2017}.
 We have tried both one-length-scale and two-length-scale models with
 the estimated parameters for the Frank-Kasper A15 structure.
 All the results shown in Fig.~\ref{fig_3Ddodecagonal} are by the
 two-length-scale model.
 The one-length-scale model cannot reproduce the quasi-crystal.
 This is natural because more than two length scales are necessary to
 form QCs.

 Using the obtained stable DDQC, we also study their fluctuations under
 the fixed amplitude of the noise.
 Figure~\ref{fig_3Ddodecagonal}(d) demonstrates the reconstruction of
 a dodecagonal ring.
 This process is due to the change of the position of
 the centre along the $c$-axis to the ring
 (see the insets of Fig.~\ref{fig_3Ddodecagonal}(d)).
 This reconstruction has also been observed in a particle-based
 model\cite{Dzugutov:1995}.
 The fluctuation is characterised by the intermediate scattering
 function, $F({\bf k},t) = \mean{\psi_{\bf k}(0) \psi^{\dagger}_{\bf k}(t)}$ for
 the wave vectors corresponding to the peaks in the Fourier space
 (Fig.\ref{fig_3Ddodecagonal}(e)).
 Here, $\dagger$ indicated complex conjugate.
 The relaxation time of $F({\bf k},t)$ indicates the diffusive time scale of
 the structure at ${\bf k}$ (Fig.~\ref{fig_3Ddodecagonal}(f)).
 The peaks other than the main peaks of the DDQC shown in
 Fig.~\ref{fig_3Ddodecagonal}(c) decay quickly in a short time scale of $t \sim 10^3$.
 Among the main peaks, the larger peaks (cyan in Fig.~\ref{fig_3Ddodecagonal}(e)) are stable against noise even in
 the longer time scale of $t \gtrsim 10^4$,
 whereas the smaller peaks (magenta in Fig.~\ref{fig_3Ddodecagonal}(e)) show diffusive behaviour.
 This slower diffusion at $t \gtrsim 10^4$ corresponds to phason flips.
 In fact, from the inverse Fourier transformation, we found the peaks with the
 intermediate amplitude (magenta in Fig.~\ref{fig_3Ddodecagonal}(e))
 correspond to the centres of the dodecagonal rings.
 \NYYY{
We stress that these analyses of fluctuations are achieved by the
 estimation of the {\it dynamical} PDE from the stationary target
 pattern in the previous section.
 }
 }
 
 \section*{Discussions and Conclusion}

In summary, we propose the inverse problem of equation discovery for a
stationary pattern \NY{ without ground truth.
We successfully estimate the best PDE for complex patterns.
}
\NY{
Our approach has several key features compared with the previous
inference methods; it is designed for the data of a stationary pattern
without ground truth,
In addition, the method has generality to apply a wide class of PDE
models and target patterns.
 Here, we summarise these features.

 {\it The estimation for stationary data: }
 \NYYY{
The previous approaches to estimate the PDE from data are estimation
 of {\it dynamical equations} from {\it dynamical non-stationary
 data}.
 In this case, the data has information about trajectories of dynamics.
In this work, we may estimate {\it dynamical equations} from {\it
 stationary data}, which is one snapshot of the stationary pattern.
 }
 Such estimation looks impossible for two reasons:
 One is lack of information of transient dynamics.
 The second reason is that in pattern formation, the
 dynamical equation is nonlinear, and therefore, $f(\psi)=0$ has
 multiple stationary solutions.
 Only one solution from the multiple solutions
 is not enough to reconstruct $f(\psi)$ uniquely.
We take the inverse problem for the stationary pattern as an estimation
 of a {\it dynamical} PDE from marginalised initial conditions (see
 Supplementary Sec.\ref{SM.sec.relatedworks.inverse.problem} \cite{seeSM} for
 comparison to other methods).
 The stable structure should be generated from a wide range of initial conditions.
In the BM-PDE, the estimation of the model reproducing a target pattern
 as a stable solution by marginalising the initial conditions as
 in equation (\ref{marginalisation.IC}) (see also
 Fig.~\ref{fig_otherfamily}(j) for demonstration).
 This is in contrast with the method to estimate time-series data, for
 example, using data assimilation, in which a single initial condition is estimated
 from data\cite{Evensen:2009}.

The order parameters play an important role in the BM-PDE. 
The order parameters can extract symmetries of the pattern and identify
the two patterns that are generated with the same parameters but
the different initial conditions.
This identification is a necessary step to marginalise the initial
patterns.


{\it The estimation without ground truth: }
The BM-PDE not only works for a numerically produced pattern, which has
ground truth of parameters, but also for an synthesised
pattern by superposition of plane waves of equation (\ref{phi.cos}).
In the latter case, our family of models does not include ground truth.
To demonstrate the ability to estimate the model for the structure
without ground truth, we have used experimental data (Fig.\ref{fig_notruth}).
Even in this case, we have successfully estimated the best PDE model and
its parameters.
This is because the BM-PDE quantifies the uncertainty by estimating the optimal noise amplitude, and is
robust against noise.

Our method may be used to find a structure that we do not know how to
reproduce by a PDE model.
We have found the three-dimensional DDQC from the
estimated parameters for its approximant.
 To our knowledge, this is the first three-dimensional DDQC found in the PDE model.

{\it The generality of BM-PDE: }
In the BM-PDE, the choice of models and the
cost function are independent.
Therefore, we may replace the models with any other PDE models.
To demonstrate this feature, we have shown the estimation of parameters
not only in the linear operator but also in nonlinear terms
(Supplementary Sec.~\ref{sec.family.Hermite} and
\ref{sec.family.nonlinear} \cite{seeSM}).
The choice of the target pattern is also flexible.
 By comparing two patterns in the space of the order parameter, we can
 estimate the model that can reproduce a similar pattern in the sense
 that it has the same symmetry to the target pattern.
 Thanks to this feature, the BM-PDE works for the noisy target.

}


\newpage

 \section*{Methods}

The main framework of BM-PDE consists of three parts: \NYY{a family of
PDE models, characterisation of a structure with order parameter, and statistical inference.
Here, we summarise their basic steps.}

\subsection*{PDE models}

We consider a pattern described by a scalar density field $\psi({\bf x})$ in a
box with a periodic boundary condition, and ${\bf x} \in [-L_x/2,L_x/2 ]
\times [-L_y/2,L_y/2 ]$ in two dimensions and ${\bf x} \in
[-L_x/2,L_x/2 ] \times [-L_y/2,L_y/2 ] \times [-L_z/2,L_z/2 ]$ in three dimensions.
The density field $\psi({\bf x})$ is obtained by an unknown model of a
partial differential equation.
In this study, we focus on a family of the phase-field crystal (PFC)
models given by
 \begin{align}
  \partial_t \psi
  &=
  \mathcal{L} \psi
  +
  \Delta  \psi^3
  \label{SHmultil}
 \end{align}
 with the linear operator denoted by $\mathcal{L}$ and the nonlinear
 term is given in the
 second term.
 The family is constructed by modifying the linear operator
 $\mathcal{L}$ so that the system has one or more length scales.
The simplest PFC model is a conserved version of the Swift-Hohenberg
 (SH) equation \cite{Provatas:2011}.
The equation is given by
 \begin{align}
  \partial_t \psi
  &=
  \Delta
  \left[
-  \epsilon \psi
  + \left(
q_0 + \Delta
  \right)^2 \psi
  + \psi^3
  \right]
  \label{SH}
 \end{align}
 where the total mass is conserved as
 \begin{align}
  \bar{\psi}
  &=
  \frac{1}{V}
  \int \psi ({\bf x}) d{\bf x}
  .
 \end{align}
Here, $V$ is the total volume (area in two dimensions) in the system,
 and $q_0$ is a characteristic wavenumber corresponding to the length
 scale $2\pi/q_0$.
 The parameter $\epsilon$ controls whether the uniform state $\psi({\bf x})
 =\bar{\psi}$ is stable $\epsilon \lesssim 0$ or unstable $\epsilon
 \gtrsim 0$.
 The precise value of the threshold is dependent on other parameters and
 a type of patterns.
 The PFC equation reproduces a stripe (also called lamellar or smectic) and
hexagonal patterns in two dimensions\cite{Pismen:2006,Elder:2004}, and a
 lamellar, hexagonal cylinder,
BCC, and hexagonal closed packing patterns \cite{Zhang:2008,Jaatinen:2010}.
A finite mean density $\bar{\psi}$ plays a role as the quadratic
  nonlinear term in equation (\ref{SH}).
  This can be seen by subtracting the mean density as $\psi \rightarrow
 \psi + \bar{\psi}$ in equation (\ref{SH}) as,
 \begin{align}
  \partial_t \psi
  &=
  \Delta
  \left[
  \left(
  -  \epsilon 
  + 3 \bar{\psi}^2
  \right)
  \psi
  + \left(
q_0 + \Delta
  \right)^2 \psi
  + 3 \bar{\psi} \psi^2
  + \psi^3
  \right]
  \label{SH2}
  .
 \end{align}
 
The PFC equation has a single characteristic length at which a real part
of the eigenvalue is positive (or, at least, negative but close to
zero).
The linear spectrum is shown in Fig.~\ref{SMspectrum}.
 To extend equation (\ref{SH}) for the arbitrary number of length
 scales, we use $m$ characteristic wavenumbers, $q_0, q_1, \cdots,
q_{m-1}$, and the values of the spectral, $a_0, a_1, \cdots, a_{m-1}$,
at the wavenumber $k=q_i$.
A family of our models is conveniently described by the equations for
the Fourier transform of the density field, that is, $\tilde{\psi} ({\bf
k}) = \mathcal{F} [\psi({\bf x})]$.
Our models corresponding to equation (\ref{SHmultil}) are given by 
\begin{align}
 \partial_t \tilde{\psi} ({\bf k})
 &=\mathcal{L}_k \tilde{\psi} ({\bf k})
 + \mathcal{F} \left[ \Delta \psi({\bf x})^3\right]
 ,
 \label{PDE.eq.FT}
\end{align}
and its linear operator is expressed in the Fourier space as
 \begin{align}
  \mathcal{L}_k
  &=
  a_0 S_0(k)
  + a_1 S_1(k)
  + \cdots
  + \NY{a_{m-1} S_{m-1}(k)}
  + 
  k^2 \left(
\NY{q_0}^2-k^2
  \right)^2
  \left(
q_1^2 - k^2
  \right)^2
  \cdots
  \left(
\NY{q_{m-1}^2} - k^2
  \right)^2
  .
  \label{linearFourier}
 \end{align}
The function $S_i (k)$ for $i \in [0,m-1]$ is
 chosen so that the coefficient $a_i$ corresponds to the peak as a
 function of $k$ of
 $\mathcal{L}_k$
 \NY{
 at $k=q_i$ (see Fig.~\ref{fig2D}(f-h)).
 Since we may freely choose a unit of length scale, we fix to be $q_0=1$
 when $i \geq 2$ in $m=m_i$.
 }
 This implies that we impose the length scale $2\pi/q_i$.
 The \NY{concrete} form of $S_i (k)$ is shown in Supplementary
 Information Sec.\ref{SM.sec.PDE.models} \cite{seeSM}.
 The parameter $s_0$ describes the sharpness of the peaks.
To make the spectrum sharp enough, we chose the parameter to be $s_0=100$ for the one-length and
	     two-length models, and
 $s_0=2000$ for the three-length model.  
 Both $a_i$ and $q_i$ are chosen as parameters to be optimised.
 We have other parameters such as the system size $L$ in each direction and
 the mean density of a pattern $\bar{\psi}$.
 In this study, we use the periodic boundary condition.
 A set of parameters is thus $\mu = \{ L_{\alpha}, \bar{\psi},
 a_0, q_0, \ldots, a_{l-1}, q_{l-1} \}$ with $\alpha \in [ 1, \ldots, d
 ]$ in space dimension $d$.

In order to make a pattern, higher-order spatial derivatives are
necessary.
 Polynomial expansion in terms of the wavenumber $k$ instead of
 equation (\ref{linearFourier}) may be available to make several
 length scales (see Fig.~\ref{fig_schematic}(e)).
 Nevertheless, it suffers from the large value of the coefficient of each
 order in the polynomials because we may not use prior distribution to
 confine the parameter space (see \NY{also} Supplementary
 Sec.~\ref{SM.sec.polynomial.expansion} \cite{seeSM}).

 \subsubsection*{Numerical simulations of PDE models}

 Numerical simulations of the PDEs are performed using the pseudo-spectral
 method in which the linear terms are computed in the Fourier space, and the
 nonlinear terms are computed in real space.
 Since our PDEs contain higher-order derivatives, we use
 the operator-splitting method\cite{Cox:2002,Aranson:2015}.
Both real and Fourier spaces are discretised into $N^d$ meshes in
 $d$-dimensional space.
 Instead of changing the system size $L_i$ in each dimension $i \in
 [1,d]$ under the periodic domain, we change the mesh size $d x_i$ so that the system size
 becomes $L_i = (N-1) d x_i$.

 The number of mesh points is fixed to be $N=128$ in two dimensions and
 $N=32$ in three dimensions.
 The larger $N$ is better in terms of accuracy, but computational time
 is scaled roughly as $\mathcal{O}(N^d)$.
 In REMC, we need to simulate it in $N_{\rm rep}$ replicas, and therefore,
 the choice of $N$ is made by the balance between accuracy and realistic
 computational time.
 In addition, the larger system size suffers from longer relaxation time and
 a higher probability that topological defects such that dislocations and
 disclinations appear.
 We set the total number of steps to be $10^5$ with a time step $d
 t=0.01$.
 We \NY{confirmed} this is enough to obtain the stationary patterns studied
 in this work, but may be changed depending on the pattern of interest.
Note that statistical inference in this study is \NY{entirely independent of}
 the algorithm of numerical simulations solving a PDE.
 An efficient algorithm would improve the performance of estimation, and
 one may replace the numerical scheme suit for one's purpose.

\subsubsection*{Formation of three-dimensional DDQC}

\NYY{
 To generate the DDQC in three dimensions, we have to add noise in
 equation (\ref{eq.PDE}),
\begin{align}
 \partial_t \psi({\bf x},t)
 &=
 \mathcal{L}_{\hat{\mu}}^{(m)} \psi ({\bf x},t) + \mathcal{N}[\psi ({\bf
 x},t)]
 + \gamma(t) \xi ({\bf x},t)
 .
 \label{SH.noise}
\end{align}
\NYY{
Note that the noise is state noise, and nothing to with the observation
noise.
}
In the annealing process, the amplitude of the noise $\gamma(t)$ is decreased in time.
The annealing schedule is chosen as $\gamma (t) = 0.198 /\log (t+\tau_a)$ with
$\tau_a=10^4$.
This choice ensures $\gamma (t) \sim 1/ \log t$ for large $t$
\cite{Geman:1984}.
The Gaussian white noise with the zero mean $\mean{\xi({\bf x},t)}=0$ and
its variance
 \begin{align}
  \mean{\xi({\bf x},t) \xi({\bf x}',t') }
  &=
-  2 \Delta \delta({\bf x}-{\bf x}') \delta (t-t')
  \label{SPDE.noise.variance}
 \end{align}
 is used.
 The Laplacian in equation (\ref{SPDE.noise.variance}) ensures the conservation
 of the density and ensures to reach the equilibrium
 state.
 We may also use other statistical properties of the noise as long as the density
 conservation is ensured.
For example, we have tested $\mean{\xi({\bf x},t) \xi({\bf x}',t') }= 2 \delta({\bf x}-{\bf x}') \delta (t-t')
 $ with $\int \xi({\bf x}',t') d {\bf x}=0$.
   We have confirmed in all cases the DDQC appears when the amplitude
   of the noise $\gamma$ is decreased.

   Fluctuation of the DDQC is studied by fixed $\gamma$.
   We set it to be $\gamma=0.171$.
    Note that the DDQC is destroyed for the noise with the amplitude of
 $\gamma^2 = 0.198$.

}

  \subsection*{Target patterns and characterisation of patterns}

\subsubsection*{Target Pattern}

A target pattern $\psi^*({\bf x})$ is prepared in two ways.
One is a numerical solution of equation (\ref{eq.PDE}) for given parameters under
a given model.
The resulting pattern is numerically transformed into the Fourier space,
and then the order parameter $\boldmath{\Psi}^*_l$ is calculated.

The second way is completely independent of the models.
 The target pattern is expressed as a density field $\psi^*({\bf x})$ that is
 a superposition of the cosine function.
 The simplest case is a stripe pattern, which is described
only by one wave
in one direction in a two-dimensional space.
Similarly, a hexagonal pattern is expressed by two-dimensional waves in
three directions.
The target pattern is expressed as
\begin{align}
 \psi^* ({\bf x})
 &=
 \sum_{i} b_i \cos ({\bf q}^*_i \cdot
 {\bf x})
 \label{phi.cos}
\end{align}
where the wave vectors ${\bf q}^*_i$ are chosen at the
position appropriate to express symmetries of the target pattern.
The amplitude of each mode $b_i$ is also chosen properly.
We numerically make the Fourier transform of equation (\ref{phi.cos}) to obtain
$\hat{\psi}^*({\bf k})=\mathcal{F}[\psi^*(\bf x)]$ and calculate the Fourier spectrum $|\hat{\psi}^*({\bf k})|$ from
which we obtain the order parameter.
The Fourier transform of equation (\ref{phi.cos}) defined in the infinite
domain is expressed by the superposition of the delta function at the
position of ${\bf q}^*_i$.
Nevertheless, the numerical Fourier transform in the bounded domain
results in peaks smeared around ${\bf q}^*_i$.
To remove the artefact, we set $|\psi^*({\bf k})|=0$ except
for the region $|\psi^*({\bf k})| > \alpha {\rm max} |\psi^*({\bf
k})|$.
Here, the value of $\alpha$ is chosen so that the peaks of the minimal height
are left.
We choose $\alpha=0.6$ for the two-dimensional target patterns, whereas $\alpha=0.01$
for the three-dimensional target patterns.

The dodecagonal QC pattern \NY{in two dimensions} is synthesised by $\psi = \sum_{i=1}^{12} \cos ({\bf q}^*_i \cdot
{\bf x})$ in which the wave vectors ${\bf q}^*_i$ are chosen at the
position of the vertices of the hexagon with a radius $|{\bf q}^*_1|=2\pi/\sqrt{2+\sqrt{3}}$ and
the hexagon with a radius $|{\bf q}^*_2|=2\pi$ rotated by $\pi/12$.
The DG pattern is expressed by 24 wave vectors of $ {\bf q}^*=(\pm 2, \pm 1,
\pm 1)$ and 12 wave vectors of $ {\bf q}^*=(\pm 2, \pm 2, 0)$ with their
permutation along the $x,y,z$ directions\cite{Schnering:1991,Yamada:2004}.
The amplitude of the latter wave vector is $\sqrt{8/6}\simeq1.15$ times
longer than the former waves.
  The FKA15 pattern is expressed by 24 wave vectors $ {\bf q}^*=(\pm 2, \pm 1, 0)$,
     24 wave vectors $ {\bf q}^*=(\pm 2, \pm 1, \pm 1)$,
  6 wave vectors of $ {\bf q}^*=(\pm 2, 0, 0)$
  with their
permutation along the $x,y,z$ directions\cite{ImperorClerc:2012}.

\subsubsection*{order parameters}
To assure translational invariance, we use a spectrum of the Fourier transform of the
 pattern and expand it by the basis functions, each of which \NY{expresses}
 certain point group symmetries.
 In two dimensions, the basis functions reflect $n$-fold rotational
 symmetry, as shown in Fig.~\ref{fig_schematic}(b), whereas in three
 dimensions, spherical harmonics expansion is used.
 Projection of the Fourier spectrum of the pattern onto the basis
 function is given by $A_{l,\pm}$ in two dimensions, and $A_{lm}$ in
 three dimensions.

 The order parameter is a rotational invariant form of the quantity
 $A_{lm}$ with $l \in [0,l_0]$ and $m \in \{ \pm 1 \}$ in two dimensions and $m
 \in [-l,l]$ in three dimensions.
In two
dimensions, $A_{l, \pm 1}$ is described by
\begin{align}
 A_{l, \pm 1} \left[ \psi \right]
 &=
 \int  |\psi({\bf k})|
\begin{pmatrix}
 \cos l \theta_k \\
 \sin l \theta_k 
\end{pmatrix}
 d{\bf k}^2
\end{align}
where $+1$ $(-1)$ corresponds to $\cos l \theta_k$ $(\sin l \theta_k)$,
respectively, and $\theta_k$ is a polar angle in the Fourier space.
We use the Fourier transform of the pattern as
\begin{align}
 \tilde{\psi} ({\bf k})
 &=
 \int
 \psi({\bf x})
 e^{i {\bf k}\cdot {\bf x}}
 d{\bf x}
 \\
  \psi({\bf x})
 &=
 \int_{\bf k}
  \tilde{\psi} ({\bf k})
 e^{-i {\bf k}\cdot {\bf x}}
\end{align}
where the volume in the Fourier space is $ \int_{\bf k} = \frac{1}{(2\pi)^d} d^d {\bf k}$.
We denote $A_{l,\pm}$ in the vector form as ${\bf
A}_l =(A_{l,+}, A_{l,-})$.
The maximum mode is denoted by $l_0$.
In three dimensions, $A_{lm}$ is given by
\begin{align}
 A_{lm} \left[ \psi \right]
 &=
 \int   |\psi({\bf k})| Y_l^m (\theta_k, \varphi_k)
 d{\bf k}^3
\end{align}
with spherical harmonics $Y_l^m (\theta_k,\varphi_k)$ in the spherical
coordinates of the Fourier space $(k,\theta_k,\varphi_k)$.
Note that $m$ in the superscript of $Y_l^m (\theta_k,\varphi_k)$ and
subscript of $A_{lm}$ should not be confused by $m$ describing a model in $M$.
 The zeroth mode $l=0$ corresponds to \NY{the mean amplitude} of
$\tilde{\psi}({\bf k})$, which is independently considered by $\bar{\psi}$.
We, therefore, use the sum for $l \in [1,l_0]$ in the cost function.
The maximum mode is denoted by $l_0$.
 We use the convention of spherical harmonics
 \begin{align}
  Y_l^m (\theta,\varphi)
  &=
  \sqrt{\frac{(2l+1)(l-m)!}{4 \pi (l+m)!}}
  P_l^m (\cos \theta) e^{i m \varphi}
 \end{align}
where $P_l^m(\cos \theta)$ is associated Legendre polynomial with
integers $l$ and $m \in [-l,l]$.
Any continuous function on a unit sphere may be expanded.

 We define the order parameter \NYY{${\bf \Psi} [\psi({\bf x})] = \{
 \Psi_l [\psi({\bf x})] \}_{l=1}^{l_0}$} by a rotationally invariant form of
 the coefficients, $A_{l,\pm}$ or $A_{lm}$.
 \begin{align}
  \Psi_l
  =
   \| {\bf A}_l\|
 &\equiv
  \sqrt{A_{l,+1}^2 + A_{l,-1}^2}
  \label{orderparameter.2D}
 \end{align}
in two dimensions, and
 \begin{align}
  \Psi_l
  =
 \| A_{lm}\|
 &\equiv
  \sqrt{\frac{4\pi}{2l+1}}
  \sqrt{
  \sum_{m=-l}^l
(-1)^m  A_{l,m} A_{l,-m}
  }
  \label{orderparameter.3D}
 \end{align}
 in three dimensions.
  Here, the prefactor is included because $ (Y_l^0)^2
+ \sum_{m=1}^l Y_l^m(\theta,\varphi) Y_l^{m*}(\theta,\varphi) =
 (2l+1)/(4\pi)$, and the sum of $|A_{lm}|^2$ scales $2l+1$.

 We numerically evaluate $A_l$ for patterns $\psi(\bf x)$ and $A^*_l$ for a
target pattern $\psi^*({\bf x})$.
Since both real space and Fourier space density fields are expressed by
values at a finite number of mesh points, the range of a mode $l$ is
truncated at the maximum mode $l_0$.
The larger mode extracts a finer structure in the Fourier spectrum, and the structure finer than the mesh size is invalid.
We thus take $l_0 = N$.
Note that for odd $l$, $A_{lm} \simeq 0$ and therefore the dimension of
$\bm{\Psi}$ is $l_0/2$.

\subsection*{Statistical inference}
\label{sec.model.selection}

\subsubsection*{Bayesian formulation}
We may extend our model naturally toward the Bayesian
formulation, which enables us not only to choose the optimal PDE, i.e. parameters and the number of the characteristic length scales, but also to evaluate their uncertainty.
To do this, we assume the order parameter \NYY{${\bf \Psi}[\psi_s]$} of the
stationary pattern $\psi_s$ is
observed as that of the target pattern $\psi^*$ with the additive noise $\xi_l$:
\begin{align}
\Psi_l^* &= \Psi_l (\psi_s; \psi_0, \mu) + \xi_l,
\label{eq:observation}
\end{align}
where $\xi_l$ is the random variable \NYY{for each mode $l$} distributed according to
zero-mean Gaussian distribution with variance $\beta^{-1} \geq 0$.
The noise $\xi_l$ and the corresponding inverse temperature $\beta$ play
a role of the uncertainty of the measurement.  
Here, $\psi_s$ and $\psi_0$ are the stationary and initial states
of $\psi$ in equation \eqref{SHmultil}, respectively, and $\mu$ is the
set of parameters.
The dependence of $\Psi_l[\psi_s]$ on $\psi_0$ and $\mu$  is explicitly represented by $\Psi_l (\psi_s; \psi_0, \mu)$.
The assumption equation \eqref{eq:observation} is equivalently represented as the conditional probability density
\begin{align}
 p(\Psi_l^* \mid \psi_0, \mu, \beta)
 &= 
 \sqrt{\frac{\beta}{2 \pi}}
 \exp \left \{ - \frac{\beta}{2} [\Psi_l^* - \Psi_l (\psi_s;  \psi_0, \mu)]^2 \right \}.
 \label{eq:likelihood}
\end{align}

We consider the parameter estimation of $\mu$ by marginalising
$\psi_0$. Hereafter,  the discretization $\psi_0 = \{ \psi_0^{(j)}
\}_{j=1}^{N^d}$ and the reparametrization $\psi_0^{(j)} \rightarrow
\bar{\psi} + \psi_0^{(j)}$ are also considered, where $\bar{\psi}$ and
$\psi_0^{(j)}$ are the mean density and the (relative) density at the
mesh point $j$\NY{, respectively}. 
By Bayes' theorem, the conditional joint probability density of $\psi_0$
and $\mu$
under given $\{ \Psi_l^* \}_{l=1}^{l_0}$, $\beta$ and the model
class $m$ is represented as
\begin{align}
 p(\mu \mid \{ \Psi_l^* \}_{l=1}^{l_0} , \psi_0, \beta, m)
 &=
 \frac{p(\mu \mid m)}
  {p(\{ \Psi_l \}_{l=1}^{l_0} \mid \beta, m)} \prod_{l=1}^{l_0} p(\Psi_l^* \mid  \psi_0, \mu, \beta) , \notag \\
 & \propto 
 \exp \left [ - \frac{\beta}{2} E( \psi^*,\psi_s;  \psi_0, \mu) \right ] 
 \label{eq:posterior}
\end{align}
where $m$ denotes the number of the characteristic length scales such as equations (\ref{SHK1}), (\ref{SHK2}) or (\ref{SHK3}).
Here, $p(\mu \mid m)$ and $p(\psi_0)$ are the {\it prior} distribution
defined as the uniform distribution, and the marginal likelihood $p(\{
\Psi_l \}_{l=1}^{l_0} \mid \beta, m)$ is given by
\begin{align}
 p(\{ \Psi_l \}_{l=1}^{l_0} \mid \beta, m)
 &= \left (
 \frac{\beta}{2 \pi}
 \right )^\frac{l_0}{2}
  \int \exp \left [ - \frac{\beta}{2} E( \psi^*,\psi_s; \psi_0, \mu)
 \right ]
 p(\psi_0) p(\mu \mid m)  d  \psi_0 d \mu .
\end{align}
The dependence of $E[ \psi^*,\psi_s]$ on $\psi_0$ and $\mu$ is explicitly represented by $E( \psi^*,\psi_s; \mu, \psi_0)$.
Note that we assume the {\it causality}
\begin{align}
p(\psi_0, \mu \mid \{ \Psi_l^* \}_{l=1}^{l_0} , \psi_0, \beta, m) = p(\psi_0) p(\mu \mid \{ \Psi_l^* \}_{l=1}^{l_0} , \psi_0, \beta, m),
\label{eq:JointProb}
\end{align}
which ignores (i) the dependence of $\psi_0$ on $\{ \Psi_l^* \}_{l=1}^{l_0}$ and $\beta$, and (ii) the correlation between $\psi_0$ and $\mu$.
This assumption reflects our ansatz that $\psi_0$ is not uniquely determined only by $\psi^*$ (or $\{ \Psi_l^* \}_{l=1}^{l_0}$).
Here $\psi_0$ is treated as a latent variable. By marginalising out $\psi_0$, the {\it posterior} distribution of $\mu$ is given by
\begin{align}
 p(\mu \mid \{ \Psi_l^* \}_{l=1}^{l_0} , \beta, m)
 &=
\int  p(\psi_0, \mu \mid \{ \Psi_l^* \}_{l=1}^{l_0} , \psi_0, \beta, m) d \psi_0.
\end{align}
The posterior mean estimator $\hat{\mu}$, i.e. the mean of $p(\mu \mid \{ \Psi_l^* \}_{l=1}^{l_0} , \beta, m)$, is adopted as our best parameter set. The standard deviation of $p(\mu \mid \{ \Psi_l^* \}_{l=1}^{l_0} , \beta, m)$ plays a role of the error in $\hat{\mu}$.

We consider both the hyperparameter estimation of $\beta$ and model
selection of $m$ \cite{efron1973stein,akaike1998likelihood,mackay1992bayesian,Bishop:2006,Tokuda:2017}.
By Bayes' theorem, the joint probability density of $\beta$ and $m$
under given $\{ \Psi_l^* \}_{l=1}^{l_0}$ is represented as
\begin{align}
p(\beta, m \mid \{ \Psi_l^* \}_{l=1}^{l_0}) &= \frac{p(\{ \Psi_l^* \}_{l=1}^{l_0} \mid \beta, m) p(\beta) p(m)}{p(\{ \Psi_l^* \}_{l=1}^{l_0})},
\end{align}
where $p(\{ \Psi_l^* \}_{l=1}^{l_0})$ is the \NY{normalisation} constant.
 Here, $p(\beta)$ and $p(m)$ are the prior distributions defined as the
 uniform distribution.
 The maximum a posteriori estimator, or equivalently the empirical Bayes estimator in this setup, is adopted as the pair of our optimal model and temperature
\begin{align}
 (\hat{\beta}, \hat{m})
 &= \argmax_{\beta, m} p(\beta, m \mid \{ \Psi_l^* \}_{l=1}^{l_0}) \\
 &= \argmax_{\beta, m} p(\{ \Psi_l^* \}_{l=1}^{l_0} \mid \beta, m).
\end{align}
For convenience, the Bayes free energy $F(\beta, m) = - \log p(\{ \Psi_l
\}_{l=1}^{l_0} \mid \beta, m) $ is defined.
Using the Bayes free energy, we may see the optimal model and
temperature $(\hat{\beta},
\hat{m})$ minimise $F(\beta, m)$.
If $\partial F/ \partial \beta = 0$ is satisfied at $\beta=\hat{\beta}$, then we obtain the self-consistent equation
\begin{align}
\hat{\beta}
 &= \frac{1}{\langle E( \psi^*,\psi_s; \psi_0, \mu) \rangle_{\hat{\beta}}},
\end{align}
where
\begin{align}
\langle \cdots \rangle_{\beta} = \int (\cdots)  p(\psi_0, \mu \mid \{ \Psi_l^* \}_{l=1}^{l_0} , \beta, m) d\psi_0 d\mu .
\end{align}
By marginalising out $\beta$, we can also evaluate the uncertainty of $m$ as the probability
\begin{align}
p(m \mid \{ \Psi_l^* \}_{l=1}^{l_0}) = \int p(\beta, m \mid \{ \Psi_l^* \}_{l=1}^{l_0}) p(\beta) d \beta.
\end{align}
Note that $p(m_1 \mid \{ \Psi_l^* \}_{l=1}^{l_0})$, ..., $p(m_{i_{\rm
max}} \mid \{ \Psi_l^* \}_{l=1}^{l_0})$
demonstrate the probability of each model $m_1$, ...,  $m_{i_{\rm
max}}$, respectively, based on the observations $\{ \Psi_l^* \}_{l=1}^{l_0}$. 

\subsubsection*{Setup of \NY{a prior} distribution}
We assume no prior information about parameters and latent
variables except their range.
The prior density of each variable is defined by the continuous uniform
distribution whose support equals the domain of each variable.
The prior density of $\psi_0$ is defined by
\begin{align}
p(\psi_0) = \prod_{j=1}^{N^d} \varphi(\psi_0^{(j)}),
\label{eq:PriorPsi0}
\end{align}
where $\varphi(\psi_0^{(j)})$ is the continuous uniform distribution whose support is $\psi_0^{(j)} \in [-0.1,0.1]$.
For equation \eqref{eq:likelihood} with $m$ length scales, the set of parameters is defined by
 \begin{align}
 \mu
   &=
 \{
dx,dy, \bar{\psi}, a_0, q_1, a_1, q_2, a_2, \cdots, q_{m-1}, a_{m-1}
 \},
 \label{eq:parameters}  
 \end{align}
  where ${\rm dim} (\mu)= 2m+2$ in two dimensions.
  In three dimensions, the mesh size along the $z$-axis $dz$ is added in
  the parameters.
The prior density of $\mu$ is also defined by
\begin{align}
p(\mu \mid m)
 &=
 \varphi(dx) \varphi(dy) \varphi(\bar{\psi}) \varphi(a_0) \prod_{i=1}^{m-1}
 \varphi(a_i) \varphi(q_i),
 \label{eq:PriorMu}
 \end{align}
where $\varphi(dx)$, $\varphi(dy)$, $\varphi(\bar{\psi})$,
$\varphi(a_i)$, and $\varphi(q_i)$ are the continuous uniform
distributions, whose supports are respectively $dx \in
[1-(1/q^*N),1+(1/q^*N)]$, $dy \in
[1-(1/q^*N),1+(1/q^*N),]$, $\bar{\psi} \in [-1,0]$, $a_i \in
[-0.2,0.2]$, and $q_i \in [0,1]$.
Here, $2\pi/q^*$ is the wavelength that is used to synthesise the target
pattern.

We also assume no prior information for model and hyperparameter;
 the prior distribution of each variable is defined by the discrete
 uniform distribution.
 The prior distribution $p(\beta)$ is also defined by the discrete
 uniform distribution with $\beta \in \{ \beta_{\alpha}
 \}_{\alpha=0}^{N_{\rm rep}-1}$, where $\beta_{0}=0$ and
 \begin{align}
  \beta_{\alpha}
  &=
  10^{
  \log_{10} \beta_{\rm min}
  + \frac{\alpha-1}{N_{\rm rep}-1} \log_{10} (\beta_{\rm max}/\beta_{\rm min})
  }
  \label{SM.beta}
 \end{align}
  for $\alpha \in \{1, 2, \cdots, N_{\rm rep}-1 \}$.
  Here, we set as $N_{\rm rep}=40$, $\beta_{\rm min}= 10^{-3}$ and
  $\beta_{\rm max}= 10^2$.
  Equation (\ref{SM.beta}) means that discretization of $\beta$ is finer
  at the large $\beta$ (lower variance of noise).
 The prior distribution $p(m)$ is defined by the discrete uniform
  distribution at \NY{$m \in \{ m_i \}_{i=1,2\ldots i_{\rm max}}$}.
Each grid point $(m_i,\beta_{\alpha})$ can be regarded as the candidate of model selection equally possible in prior.
 
\subsubsection*{\NY{sampling from a posterior distribution with Monte Carlo method}}

\NY{
 The realisation of $p(\mu \mid {\bf \Psi}^*, \psi_0, m,\beta)$ is carried out by Monte Carlo (MC) sampling in the parameter space.
 For each point in the parameter space, we compute a stationary pattern $\psi_s$ in the
 model of equation (\ref{eq.PDE}) under the randomly \NYY{chosen initial condition} $\psi_0$.
 \NYY{Then, the parameters are changed} stochastically according to the
 Metropolis criterion defined by the cost function (energy) $E [\psi, \psi^*]$ and inverse temperature $\beta$. 
We use the REMC method \cite{Geyer:1991,Hukushima:1996} with different
 inverse temperatures $\beta$ in parallel.
The REMC is an efficient method for the estimation of the optimal variance $\hat{\beta}^{-1}$
 of the observation noise because the method enables us to sample parameters simultaneously under
 various $\beta$.
 The method also makes an efficient sampling to avoid a local trap in the parameter space.
Bridge sampling \cite{meng1996simulating, gelman1998simulating} is used to calculate $F(m, \beta)$ for each $m$. The error bars of $F(m, \beta)$ are calculated by the bootstrap resampling \cite{efron1992bootstrap}.
}

The joint distribution $p(\psi_0, \mu \mid \{ \Psi_l^* \}_{l=1}^{l_0} ,
\beta, m)$ is realised by the Gibbs sampling based on the relation
of equation \eqref{eq:JointProb}, i.e. the alternately iterative sampling from $p(\psi_0)$ and $p(\mu \mid \{ \Psi_l^* \}_{l=1}^{l_0} , \psi_0, \beta, m)$. 
The sampling from $p(\psi_0)$ simply follows equation \eqref{eq:PriorPsi0}. 
The sampling from $p(\mu \mid \{ \Psi_l^* \}_{l=1}^{l_0} , \psi_0, \beta, m)$ follows the procedure below.
First, we solve the model of equation \eqref{SHmultil} under a given initial state $\psi_0$ and 
parameters $\mu$ of equation \eqref{eq:parameters} for a model $m$.
Then, the similarity of an obtained pattern $\psi_s$ as a stationary
state and a target pattern $\psi^*$ is evaluated by the cost function
$E( \psi^*,\psi_s;  \psi_0, \mu)$, describing the distance shown
in equation
\eqref{cost.function} in the space of the order parameter $\Psi_l$
(see equation \eqref{orderparameter.2D} or equation \eqref{orderparameter.3D}).
Changing $\mu$, we may iterate numerical simulations and
evaluation of the similarity between them.
Following the Metropolis criterion at an inverse temperature $\beta$, we compare a current cost function with a
cost function in a previous step, and decide a current set of
parameters is accepted or not.  

By using the replica-exchange Monte Carlo (REMC) method, we sample
 $\psi_0$ and $\mu$ from $p(\psi_0, \mu \mid \{ \Psi_l^* \}_{l=1}^{l_0},
 \beta_{\alpha}, m)$ for  $N_{\rm rep}$ replicas in parallel \cite{Geyer:1991,Hukushima:1996}.
 At higher temperature \NY{(smaller $\beta$)}, the motion of one MC step in the parameter space is
 large, whereas at the lower temperature \NY{(larger $\beta$)}, each motion is small so that it
 intensively samples parameters near the minimum of the cost function.
 For every two steps, the parameter sets of neighbouring \NY{$\beta$} were
 exchanged following the Metropolis criterion.
 This process enables us to sample parameters weighed with likelihood
 effectively \cite{Geyer:1991,Hukushima:1996}.

 The initial parameter set is sampled from the prior distribution
 of equation \eqref{eq:PriorMu}.
 The lowest \NY{cost function} is typically achieved by $1000-2000$ MC steps.
 In one MC step, the Gibbs sampling is used to perform motion in the
 parameter space in all directions one by one.
 After finding the lowest \NY{cost function}, we restart REMC from the initial
 parameter set of the lowest energy state to sample its steady state.
 This is because the relaxation under \NY{smaller $\beta$} is much faster
 than the \NY{larger $\beta$}.
 After $1000$ MC steps, we cut the initial burn-in steps and compute the
 statistical quantities after $200$ MC steps. Bridge sampling was used
 to calculate $F(\beta, m)$ for each $m$ \cite{meng1996simulating, gelman1998simulating}. The error bars of $F(\beta, m)$ were calculated by the bootstrap resampling \cite{efron1992bootstrap}.

\subsection*{Regression method for noisy data}

\NYY{
In the BM-PDE, the estimation from data with noise is performed by
adding zero-mean Gaussian noise $\xi$ in the target pattern, namely $\psi^*
({\bf x}) \rightarrow \psi^* ({\bf x}) + \xi$.
In addition to this estimation using BM-PDE, another parameter
estimation method is tested.}
  We performed parameter estimation using the regression method
  in which the following cost function was used
  \NY{
\begin{align}
 E
 &=
 \frac{1}{2}
 \int \left[
 \psi^*
 - f_{\mu}^{d t}  (\psi^*)
 \right]^2 d {\bf x}
\label{nonlinear.regression}
 .
\end{align}
Here, the target pattern $\psi^*$ is numerically forward in $d t$ by
the model of equation (\ref{eq.PDE}) with a parameter set $\{ \mu \}$ as $
f_{\mu}^{d t} (\psi(t) ) = \psi(t+d t) $.
}
If the target pattern is the stationary solution of the model, \NY{namely,} if
the parameters are ground truth to obtain the target pattern, the
cost function must be zero.
This approach is philosophically the same as the regression method in previous
studies in which the cost function is a difference between \NY{the} {\it left-hand side} (time
derivative) and \NY{the} {\it right-hand side} (force to change $\psi$), that is,
\NY{$E = (1/2)\|
 \partial_t  \psi - f_{\mu}(\psi)\|$}, under an
 appropriate norm $\| \cdot \|$ \cite{Brunton:2016,Schaeffer:2017}.
 The norm is typically chosen as the $L^2$ norm
\begin{align}
 E
  &=
  \frac{1}{2} \int
  \left[\partial_t  \psi - \NY{f_{\mu}(\psi)} \right]^2 d{\bf x}
  \label{nonlinear.regression.dt}
  .
\end{align}
 In the current system, our model is no longer linear in the parameters,
 and therefore, we cannot use linear regression (including conventional
 sparse regression).
 In order to carry out nonlinear regression, we used the REMC method
 to minimise the cost function, equation (\ref{nonlinear.regression}).
The method is similar to our main algorithm to sample parameters and to
 estimate the optimal noise by temperature $\beta$.
 Following Bayes' theorem, we estimate the best parameters by the
 sampled values and their error by the standard deviation.

\newpage

\section*{Code availability}
All codes used in this work are written in MATLAB and freely available from the corresponding
author upon a request.

\section*{Data availability}
The data that support the findings of this study are availablefrom the corresponding author on reasonable request.

\begin{acknowledgments}
The authors are  grateful to Edgar Knobloch, Yasumasa Nishiura,
 An-Chuang Shi, and Philippe Marcq for
 helpful discussions.
 The authors acknowledges the support by JSPS KAKENHI grant numbers 17K05605,
 20H05259, and 20K03874 to N.Y. and 20K19889 to S.T.
  Numerical simulations in this work were carried out in part by AI
 Bridging Cloud Infrastructure (ABCI) at National Institute of Advanced Industrial Science and Technology (AIST).
\end{acknowledgments}

\section*{Author contribution}

N.Y. and S.T. conceived the research. N.Y. carried out simulations.
N.Y. and S.T. analysed the results.
N.Y. and S.T. wrote the manuscript.

\section*{Materials \& Correspondence}
Correspondence and requests for materials should be addressed to N.Y.
 \section*{Competing Interests statement}
The authors declare no competing financial interests.



\newpage

\section*{Figures and Tables}

\newpage

\newpage

\newpage

\newpage

\newpage

\end{document}